\documentclass[a4paper,11pt]{article}
\pdfoutput=1 % if your are submitting a pdflatex (i.e. if you have
\usepackage{jheppub}
\usepackage{float}
\usepackage{graphicx}
\usepackage{amsmath} 
\usepackage{ulem}

\usepackage{amsmath,graphicx,amssymb,subfigure,bbm}

\usepackage[utf8]{inputenc}
\usepackage[T1]{fontenc}

\usepackage{bbold}

\usepackage[all]{xy}

\usepackage{verbatim}

\newcommand{\be}{\begin{equation}}
\newcommand{\ee}{\end{equation}}
\newcommand{\bea}{\begin{eqnarray}}
\newcommand{\eea}{\end{eqnarray}}

\def\beq{\begin{equation}}
\def\eeq{\end{equation}}
\def\beqa{\begin{eqnarray}}
\def\eeqa{\end{eqnarray}}

\begin{document}

\title{Chiral transition and meson melting within improved holographic soft wall models}

\author[a]{Alfonso Ballon-Bayona,}
\author[b]{Sean Bartz,}
\author[c,d]{Luis A. H. Mamani,}
\author[e]{and Diego M. Rodrigues}
\affiliation[a]{Instituto de F\'{i}sica, Universidade
Federal do Rio de Janeiro, \\
Caixa Postal 68528, RJ 21941-972, Brazil.}
\affiliation[b]{
Dept.~of Chemistry and Physics, Indiana State University,\\
Terre Haute, Indiana 47809, United States of America
}
\affiliation[c]{Centro de Ci\^encias Exatas e Tecnol\'ogicas, Universidade Federal do Rec\^oncavo da Bahia,\\
Rua Rui Barbosa, 710, 44380-000, Cruz das Almas, Bahia, Brazil}
\affiliation[d]{Laborat\'orio de Astrof\'isica Te\'orica e Observacional,\\ Departamento de Ci\^encias Exatas e Tecnol\'ogicas,\\ 
Universidade Estadual de Santa Cruz, 45650-000, Ilh\'eus, Bahia, Brazil}
\affiliation[e]{Instituto de Física Teórica, UNESP-Universidade Estadual Paulista,\\ R. Dr. Bento T. Ferraz 271, Bl. II, Sao Paulo 01140-070, SP, Brazil}

\emailAdd{aballonb@if.ufrj.br}
\emailAdd{sean.bartz@indstate.edu}
\emailAdd{luis.mamani@ufrb.edu.br}
\emailAdd{diegomhrod@gmail.com}

\abstract{We describe the chiral transition for the quark condensate and the melting of scalar and vector mesons in two-flavor holographic QCD. This is done by extending  the improved holographic soft wall models proposed in \cite{Ballon-Bayona:2021ibm} to finite temperature, by means of introducing an asymptotically AdS black brane.  We find that the chiral transition is second order in the chiral limit and a crossover for physical quark masses, as expected in two-flavor QCD. We investigate the melting of vector and scalar mesons in the deconfined plasma through the calculation of  hadronic spectral functions. Fixing the model parameters by the meson spectrum at zero temperature, we find that the mesons melt at temperatures between $90$ and $110$  MeV and the chiral transition occurs around $129$ MeV. We also provide a prediction for the hydrodynamic diffusion constant associated with a flavor current in the deconfined plasma. }

\maketitle
%\flushbottom

\section{Introduction}

Describing the QCD phase diagram is one of the most important problems in high energy physics in the last decades. It allows us to extend the physics of ordinary hadrons to the realm of finite temperature and finite baryon density. One important aspect of the QCD phase diagram is the fate of chiral symmetry. At zero temperature and zero density, chiral symmetry is an approximate symmetry in the sector of light quarks that is spontaneously broken by the QCD vacuum. This is characterized by a non-zero quark condensate which in turn contributes to hadronic mass generation. At high temperatures (or high densities) it is expected that the chiral symmetry is restored, the so-called chiral transition. In the case of zero baryon density, lattice QCD results indicate that the chiral transition is a crossover with a pseudocritical temperature close to the deconfinement temperature. 

Lattice QCD has been established as the main tool to describe QCD in the strongly coupled regime and it successfully describes the chiral and deconfinement transition at finite temperature and zero baryon density. There are, however, two important limitations in the lattice approach.  One of them is the so-called sign problem at finite baryon density and the other is the description of real-time hadronic correlation functions.  Other non-perturbative approaches are needed in order to  address these issues. This includes the Schwinger-Dyson equations, Nambu-Jona-Lasinio models, chiral effective theory, and holographic QCD.  In this paper we are interested in the holographic QCD approach that allows the investigation of confinement and conformal symmetry breaking and naturally incorporates hadronic resonances. We are particularly interested in investigating the effects of conformal symmetry breaking on the chiral transition and on the real-time hadronic spectral functions.

The melting of mesons is an important observable in the quark-gluon plasma produced in heavy ion collisions. It is expected that the binding between quarks is reduced due to color screening in the deconfined plasma and as a consequence the mesons dissociate. For a review of the melting of light mesons see \cite{Rapp:2006xzj} and for the melting  of heavy mesons see \cite{Mocsy:2013syh}. The melting of light mesons in holographic QCD  was originally investigated in \cite{Hoyos-Badajoz:2006dzi,Peeters:2006iu} within the top-down approach and in \cite{Fujita:2009wc,Colangelo:2009ra} within the bottom-up approach, see also \cite{Mamani:2013ssa,Bartz:2016ufc}. The melting of heavy mesons in holographic QCD has been investigated in \cite{Grigoryan:2010pj,Dudal:2014jfa,Braga:2017bml,Jena:2024cqs}. 

In this paper we investigate the chiral transition and the effects of chiral restoration on the melting of light scalar and vector mesons. In order to describe the chiral transition we consider the improved soft wall model proposed recently in \cite{Ballon-Bayona:2021ibm} where the spontaneous breaking of chiral symmetry is described in terms of a non-linear tachyon potential and a non-mimimal dilaton coupling. This model was inspired by the previous proposals in \cite{Chelabi:2015gpc,Chelabi:2015cwn} and \cite{Ballon-Bayona:2020qpq}. 
We extend the model of \cite{Ballon-Bayona:2021ibm} to finite temperature and show that at sufficiently high temperatures chiral symmetry is restored and the pseudocritical temperature for the chiral transition is close to the lattice QCD results.  We  investigate the melting of vector and scalar mesons in holographic QCD via the dynamics of  5d field perturbations. Investigating the hydrodynamic limit of the vectorial sector we make a prediction for the diffusion constant associated with a flavor current in a deconfined plasma. The melting temperature for each hadronic sector is obtained  from the analysis of the real time spectral functions associated with the 4d meson operators.

The organization of this paper is as follows.  In section \ref{Sec:Model} we describe  the holographic model of \cite{Ballon-Bayona:2021ibm} and its extension to finite temperature. In section \ref{sec:chiral} we focus on the dynamics of the background tackyonic field leading to the chiral transition for the quark condensate. In section \ref{Sec:mesons} we describe the perturbations of the holographic model focusing on the effective Schr\"odinger potentials and we also calculate the hydrodynamic diffusion constant. In section \ref{sec:spectral_functions} we calculate the spectral functions associated with the melting of vector and scalar mesons.  We present our conclusions in section \ref{Sec:Conclusions} and additional material in appendices \ref{App:Condensate}, \ref{App:Hydro} and \ref{App:Spectral_numerics}.

\section{The holographic QCD model}
\label{Sec:Model}

In this work we  investigate the chiral transition for the quark condensate and the melting of light  scalar and vector mesons using a holographic approach to two-flavor QCD in the large $N_c$ limit. We  extend to finite temperature the improved soft wall model proposed in \cite{Ballon-Bayona:2021ibm}. 

In the improved soft wall model of \cite{Ballon-Bayona:2021ibm} the quark condensate and the chiral currents are mapped to a 5d scalar field (the tachyon) and 5d non-Abelian gauge fields respectively.  These fields in turn are minimally coupled to a 5d metric and non-minimally coupled to a background scalar field (the dilaton). The action can be written as 
\begin{align}
S &= - \int d^5 x \sqrt{-g}\,{\rm Tr}\Big \{ e^{-a(\Phi)} \Big[  |D_m X|^2 + V(|X|)   \Big]  
+ \frac{e^{- b(\Phi)}}{4 g_5^2} \,
\Big [{F_{mn}^{(L)}}^2 + {F_{mn}^{(R)}}^2 \Big ] \Big \}  \,,  \label{Eq:5dmodel}
\end{align}
\noindent
where 
\begin{align}
F_{mn}^{(L/R)} &= \partial_m A_n^{(L/R)} - \partial_n A_m^{(L/R)} - i [ A_m^{(L/R)} , A_n^{(L/R)} ]  \\
D_m X &= \partial_m X - i A_m^{(L)} X + i X A_m^{(R)} \, , 
\end{align}
and the tachyon potential is given by 
\begin{equation}
V(|X|) = m_X^2 |X|^2 + \lambda |X|^4 \,, 
\end{equation}
with $m_X^2=-3$. The dilaton couplings are $\exp(-a(\Phi))$ and $\exp(-b(\Phi))$. The dimensionless gauge coupling $g_5^2$ is fixed as usual to the value $g_5^2=12 \pi^2/N_c$ in order to reproduce the perturbative (large $N_c$) QCD result for the vectorial correlation function at small distances \cite{Erlich:2005qh}.  

At zero temperature the background is simply given by the $AdS_5$ metric \footnote{In this work we set the AdS radius to one.}
\begin{align}
ds^2 &= e^{2 A_s(z)} \left ( - dt^2 + dz^2  + d \vec{x}^2  \right )   \quad , \quad A_s(z) = - \ln z \, ,
\end{align}
and a  dilaton field quadratic in the radial coordinate
\begin{eqnarray}
\Phi = \Phi(z) = \phi_{\infty} z^2 \,.     \label{Eq:Dilaton}
\end{eqnarray}
The quadratic dilaton was first proposed in the original soft wall model  \cite{Karch:2006pv} as a mechanism for conformal symmetry breaking and  hadronic mass generation where the effective Schr\"odinger potential takes the form of a harmonic oscillator, leading to linear Regge trajectories $m_n^2 \sim n$ for large $n$, where $n$ is the radial excitation number. It was later shown that a quadratic dilaton can also be obtained in confining holographic QCD models based on Einstein-dilaton gravity \cite{Gursoy:2007er,Li:2013oda}, see also \cite{Ballon-Bayona:2017sxa,Ballon-Bayona:2024yuz}.

As explained in \cite{Ballon-Bayona:2021ibm}, the non-minimal coupling between the dilaton and the tachyon, i.e.  $\exp(-a(\Phi))$, combined with the form of the tachyon potential $V(|X|)$ induces spontaneous chiral symmetry breaking. We choose to work with model IIA of \cite{Ballon-Bayona:2021ibm} where 
\begin{eqnarray}
a(\Phi) = b(\Phi) = \Phi - a_0 \frac{ \Phi^{3/2}}{1 + \Phi^2} \quad \text{where} \quad \Phi(z) = \phi_{\infty} z^2  \,. 
\label{ModelIIA}
\end{eqnarray}
In this scenario the vectorial sector is coupled to the dilaton in the same way as the scalar sector. The parameter $a_0$ controls the breaking of conformal symmetry at intermediate energies and above some critical value induces the spontaneous breaking of chiral symmetry. Note that far from the boundary (infrared regime) the dilaton couplings become $\exp(-\Phi)$, as expected in string theory. 

In order to extend the holographic QCD model of \cite{Ballon-Bayona:2021ibm} to finite temperature, we consider the metric of a 5d asymptotically AdS black brane: 
\noindent
\begin{equation}\label{Eq:BHMetric}
ds^2=e^{2A_s(z)}\left(- f(z) dt^2+\frac{dz^2}{f(z) }+ d \vec{x}^2 \right) \, ,
\end{equation}
\noindent
where the horizon (blackening) function, $f(z)$, and the warp factor, $A_s(z)$, are given by
\noindent
\begin{equation}
f(z)=1-\frac{z^4}{z_h^4},\qquad A_s(z)=- \ln z \,,
\end{equation}
\noindent
and the dilaton is given by \eqref{Eq:Dilaton}. The radial coordinate is now defined in the range $0 < z \leq z_h$.   Note that the horizon function $f(z)$ is equal to one in the limit $z \to 0$ (near the boundary) and vanishes at the event horizon, $z=z_h$. Near the horizon the function $f(z)$ can be expanded as 
\begin{align}
f(z) = f_{h,1} (z - z_h ) + {\cal O} (z - z_h)^2 \, , \quad 
f_{h,1} = - \frac{4}{z_h} \,. 
\end{align}
The temperature of the black brane is given by \footnote{The temperature is found by imposing the absence of singularities for the near horizon metric when the time coordinate is imaginary and periodic, with period $\beta =1/T$.} 
\noindent
\begin{equation}\label{Eq:Temperature}
    T= \frac{|f'(z_h)|}{4 \pi} = - \frac{f_{h,1}}{4\pi} = \frac{1}{\pi z_h}
\end{equation}
This is also interpreted as the temperature of the dual deconfined plasma in 4d. 
The entropy density of the AdS black brane is given by the Bekenstein-Hawking area formula 
\begin{align}
 {\cal S} &= \frac{A_h}{4 G_5 V_3} = 4 \pi \sigma e^{3 A_s(z_h)} = 4 \pi \sigma z_h^{-3} \,. 
\end{align}
In this work it will be useful to define the tortoise coordinate $r^*$ by the equation 
\begin{equation}
d r^* =- \frac{dz}{f(z)} \, , \label{torteq}
\end{equation}
so that light rays in the $(t,r^*)$ have the usual form. The solution to \eqref{torteq} such that $r^*=0$ at the boundary $z=0$ is given by 
\begin{equation}
r^*(z) = \frac{z_h}{4} \Big [ -2 \arctan \left (\frac{z}{z_h} \right ) + \ln \left ( \frac{z_h - z}{z_h+z} \right )\Big ] \,.
\end{equation}
Note that the horizon $z=z_h$ corresponds to the limit $r^* \to - \infty$.

We finish this section describing the free parameters in our holographic model. The parameters $a_0$ and $\lambda$ are related to the spontaneous breaking of chiral symmetry while the parameter $\phi_{\infty}$ leads to a mass scale for the mesons.  We  fix the parameters as in \cite{Ballon-Bayona:2021ibm} to the values $a_0=6.5$, $\phi_{\infty}=(0.388 \, {\rm MeV})^2$ and $\lambda=60$, in order to reproduce the spectrum of mesons at zero temperature. The tachyon field will bring a new parameter $m_q$ associated with the physical quark mass, as described in the next section.  For the chiral transition we will consider two possible scenarios: i) $m_q=0$, which corresponds to the chiral limit and ii) $m_q = 9 \, {\rm MeV}$, which is the value used in \cite{Ballon-Bayona:2021ibm} for the physical mass of $u$ and $d$ quarks.

In the next section we will describe the dynamics of the tachyonic field $X(z)$ due to the action in \eqref{Eq:5dmodel} and the black brane background in \eqref{Eq:BHMetric}. This will naturally map to the finite temperature chiral transition of the dual 4d field theory. The melting of scalar and vector mesons will be described in sections \ref{Sec:mesons} and \ref{sec:spectral_functions} in terms of the dynamics of the perturbations of 5d scalar and gauge fields.

\section{Spontaneous chiral symmetry breaking and the chiral phase transition} \label{sec:chiral}
\noindent

In this section we first present the tachyon differential equation and its asymptotic expansions near the horizon ($z\to z_h$) and near the boundary ($z\to0$). Then, we provide details on the numerical procedure used for solving this equation and describe our numerical results for the chiral phase transition. Finally, we obtain the free energy density associated with the quark condensate and discuss the stability of the solutions at finite quark mass.

\subsection{Tachyon differential equation and asymptotic expansions}
\label{SubSec:TachyonEq}
\noindent

In two-flavor QCD the spontaneous breaking of chiral symmetry is associated with a nonzero value of the quark condensate. In holographic QCD the quark mass operator maps to the tachyonic scalar field $X$, so we consider the following ansatz:
\begin{equation}
A_{m}^{(L/R)} = 0 \quad, \quad 
X(z) = \frac12 \chi(z) {I}_{2 \times 2} \,.
\end{equation}
Under this ansatz the action in \eqref{Eq:5dmodel} takes the 1d form
\begin{align}
S_0 &=  - V_4 \int dz \, e^{3A_s - a} \Big [ \frac12 f (\partial_z \chi)^2 + e^{2A_s} V(\chi)  \Big ] \, ,
\end{align}
where $V_4 = \int d^4 x $ and the tachyon potential becomes 
\begin{equation}
V(\chi) = \frac{m_X^2}{2} \chi^2 + \frac{\lambda}{8} \chi^4 \, , \label{Vpot}
\end{equation}
with $m_X^2=-3$. 
The differential equation for the tachyon takes the form
\begin{equation}
\label{tachyoneq}
  \chi''+ \left(3A_{s}' - a' + \dfrac{f'}{f}  \right)\chi' -\dfrac{e^{2A_s}}{f}\partial_{\chi}V(\chi) = 0 \,.
\end{equation}
We remind the reader that $A_s(z) = -\ln z$ is the AdS warp factor, $f(z) = 1-z^4/z_h^4$ is the horizon function, such that $f(z_{h})=0$,  and $a(\Phi)$ is the dilaton coupling given  in \eqref{ModelIIA}. Note that $a'$ refers to the total derivative of the function $a\Phi(z))$ with respect to the radial coordinate $z$. 

Since the tachyon equation \eqref{tachyoneq} is a second-order nonlinear ordinary differential equation, one must impose two boundary conditions to specify a solution. Near the horizon (IR), we choose the solution that is regular around the horizon $z=z_h$ and therefore can be Taylor expanded as
\begin{equation}
\chi(z\to z_{h})  = C_0 + C_1 (z-z_h) + C_2 (z-z_h)^2+...
\end{equation}
Plugging this expansion into \eqref{tachyoneq} and solving for the coefficients we find
\begin{equation}\label{irexp}
\chi(z\to z_{h}) = C_0 + V'(C_0) \Big [ \dfrac{1}{4z_{h}}(z-z_{h}) - \dfrac{(-8+4z_{h}a'(z_h)-V''(C_0))}{64z_{h}}(z-z_{h})^2+...\Big ].
\end{equation}
Note that, in the near-horizon expansion for the tachyon, all orders in $\mathcal{O}(z-z_h)^{n}$ are proportional to the derivative of the potential evaluated at $C_0$. In fact, when $V'(C_0)=0$ we have an exact solution of the full tachyon equation \eqref{tachyoneq}. For the tachyon potential we are considering, we have
\begin{equation}
\label{extremaPotential}
\partial_{\chi}V(\chi) = 0 \longrightarrow \chi = \left(0,\,\pm\sqrt{\dfrac{6}{\lambda}}\right),
\end{equation}
which corresponds to the exact solutions at $T = 0$, and one can check that they are still  exact solutions of the finite temperature tachyon equation \eqref{tachyoneq}.

Near the boundary (UV), on the other hand, the tachyon behaves as
\begin{equation}
  \chi(z\to0) = m_{q}\zeta\,z + z^3 \Big [ \dfrac{\sigma}{\zeta} + d_{1} \ln ( \phi_{\infty}^{1/2} z ) \Big ]  +d_{2}\,z^4 +...\,. 
\end{equation}
where $m_q$ is identified as the current quark mass $m_u=m_d$ and $\sigma$ is identified as the quark condensate $\langle \bar u u \rangle = \langle \bar d d \rangle$. In fact, the quark mass and quark condensate can be written as $m_q = \phi_{\infty}^{1/2} c_1$ and $\sigma = \phi_{\infty}^{3/2} c_3$ where $c_1$ and $c_3$ are the dimensionless UV coefficients that characterize the model, see \cite{Ballon-Bayona:2020qpq,Ballon-Bayona:2021ibm} The constant $\zeta =\sqrt{N_c}/(2\pi)$ is fixed in order to give the correct $N_c$ scaling behavior of the quark mass and chiral condensate \cite{Cherman:2008eh}. Plugging this expansion into \eqref{tachyoneq} one can solve for the coefficients $d_{i}$, and we end up with
\begin{equation}\label{uvexp}
\chi(z\to0) = m_{q}\zeta\,z +\dfrac{\sigma}{\zeta}\,z^3 + m_{q}\left(\dfrac{1}{4}m_{q}^{2}\zeta^{3}\lambda-\zeta\phi_{\infty}\right)z^3 \ln ( \phi_{\infty}^{1/2} z ) - a_{0}m_{q}\zeta\phi_{\infty}^{3/2}\,z^4+...
\end{equation}

Finally, with both asymptotic expansions \eqref{irexp} and \eqref{uvexp} one can construct a matching routine which numerically integrates the differential equation \eqref{tachyoneq} and extracts the parameters $(C_0, \sigma)$ for a given quark mass $m_q$ as a function of temperature. 
First, we numerically integrate \eqref{tachyoneq} from the horizon ($z=z_{h}$) towards the boundary ($z=0$) till a middle point ($z_{m} =z_h/2$)\footnote{The integration range is such that $z\in(0, z_{h})$, but one can set $z_{h}=1$ by rescaling the holographic radial coordinate $z$.} using the boundary condition \eqref{irexp}. Then, we numerically integrate \eqref{tachyoneq} from the boundary ($z=0$) towards the horizon  ($z=z_{h}$) till a middle point ($z_{m} = z_h/2$) using the boundary condition \eqref{uvexp}. Finally, by matching the numerical solution for $\chi$ and its derivative $\chi'$ at the middle point, one can extract the pair ($C_0,\,\sigma$) for a given quark mass $m_q$ as a function of the temperature $T$. For a similar numerical analysis of the tachyon differential equation see, for example, \cite{Cao:2020ryx}. 

\subsection{Chiral phase transition: Numerical results} \label{subsec:chiral_numerical}
\noindent

In this subsection we present our numerical results for the tachyon profile $\chi(z)$ as a function of the holographic coordinate $z$ obtained by numerically solving \eqref{tachyoneq}. We also describe and discuss our results for the chiral phase transition in the chiral limit and for finite quark mass. Throughout this subsection we fix the dilaton constant $\phi_{\infty}$ and the quartic coupling $\lambda$ in our model as follows: $\phi_{\infty} = (388\, \mathrm{MeV})^2$ and $\lambda = 60$.  Furthermore, the results for the chiral transition are presented for two values of the dilaton coupling parameter: $a_0 = 6.5$ and $a_0 = 8.7$. The first value was found in \cite{Ballon-Bayona:2021ibm} by a fit the zero-temperature scalar meson spectra. The second value is determined by matching the latest lattice QCD result for the transition temperature in the case of physical quark masses \cite{Borsanyi2020}.

\begin{figure}[htb]
   \centering    
   \includegraphics[scale=0.37]{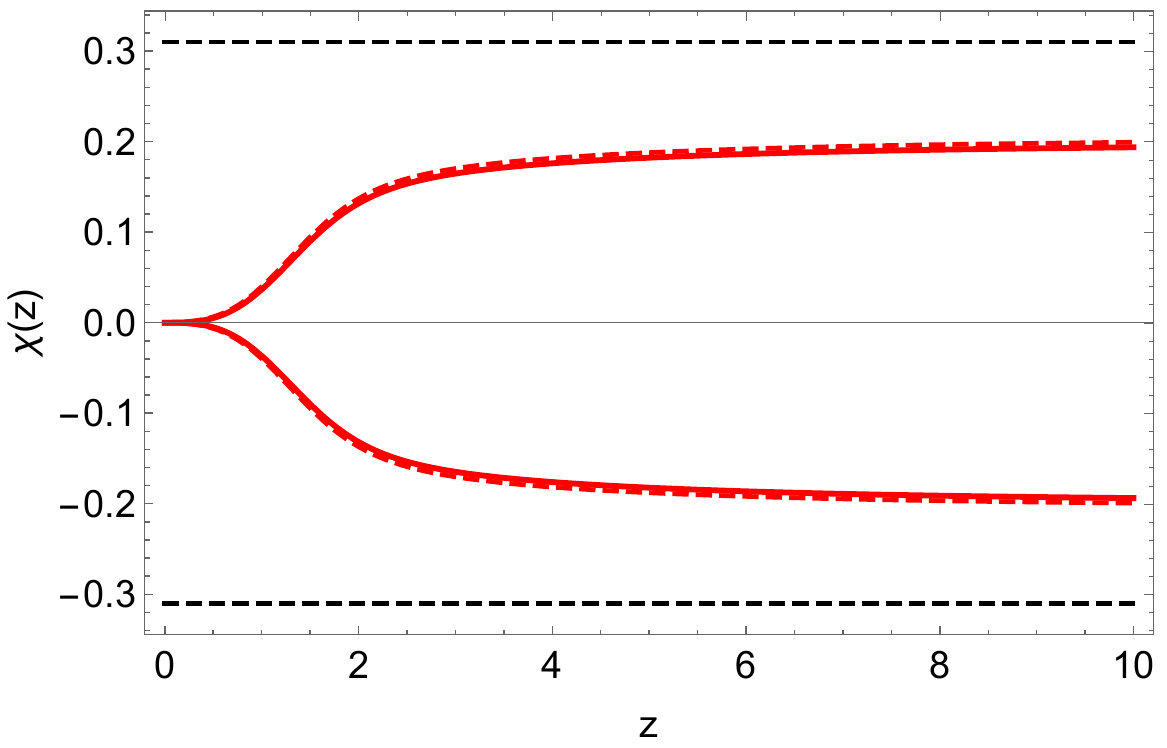}
   \hfill
   \includegraphics[scale=0.37]{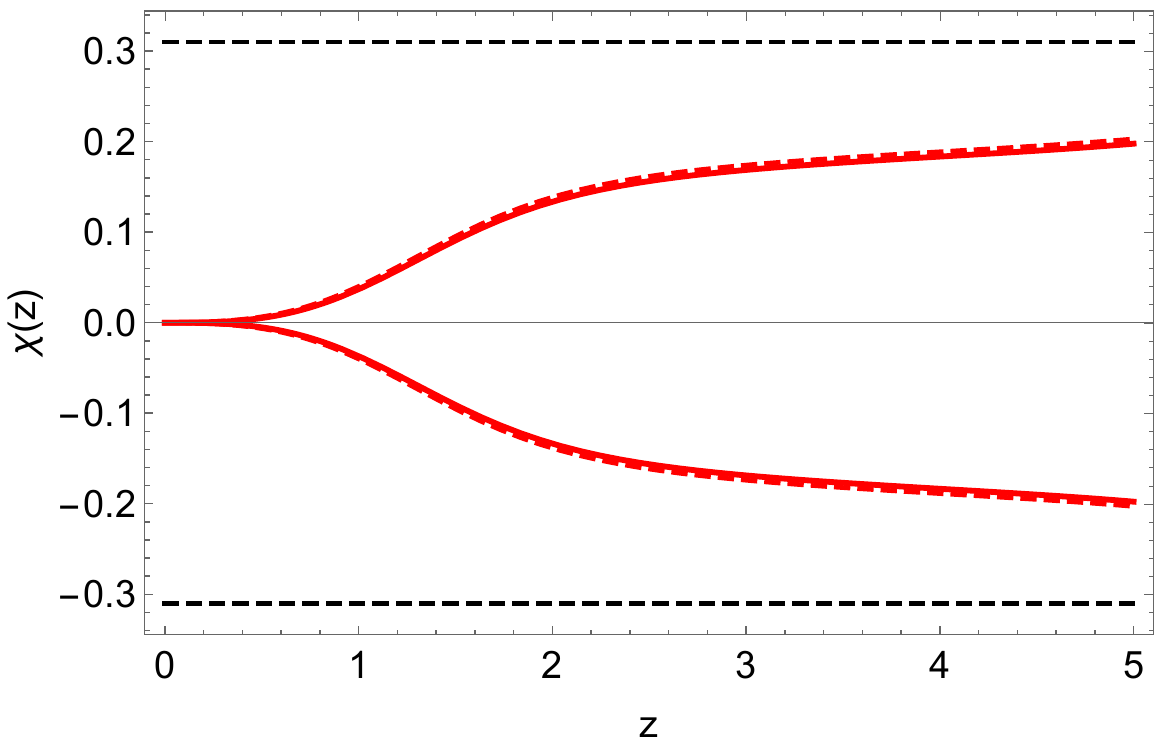}
    \caption{\textbf{Left Panel}: Tachyon profile $\chi(z)$ as a function of the holographic coordinate $z$ at zero temperature ($T=0$). \textbf{Right Panel}: Tachyon profile $\chi(z)$ as a function of the holographic coordinate $z$ at finite temperature ($T \approx 64$ MeV). In both plots the thick red curves represent the numerical solution in the chiral limit ($m_q = 0$), while the dashed red ones represent the numerical solution for finite quark mass ($m_q = 9$ MeV), and the horizontal dashed black curves represent the exact solutions, which correspond to the minima of the tachyon potential $V(\chi)$.}
\label{Fig:tachyonProfile}
\end{figure}

In Fig.~\ref{Fig:tachyonProfile} we display the typical radial profile of the tachyon field $\chi(z)$ found by numerically solving the tachyon equation \eqref{tachyoneq} at zero and finite temperatures (left and right panels, respectively). In both plots the thick red curves represent the numerical solution for $\chi(z)$ in the chiral limit ($m_q = 0$), while the dashed red ones represent the numerical solution for finite quark mass ($m_q = 9$ MeV). Note that, due to the $\mathbb{Z}_{2}$ symmetry ($\chi\leftrightarrow-\chi$) of the action, we have both the positive and negative solutions as well as the trivial one ($\chi=0$), which is not shown in Fig.~\ref{Fig:tachyonProfile}. The horizontal dashed black curves represent the nonzero constant solutions in \eqref{extremaPotential}.

\begin{figure}[htb]
\centering
\includegraphics[scale=0.37]{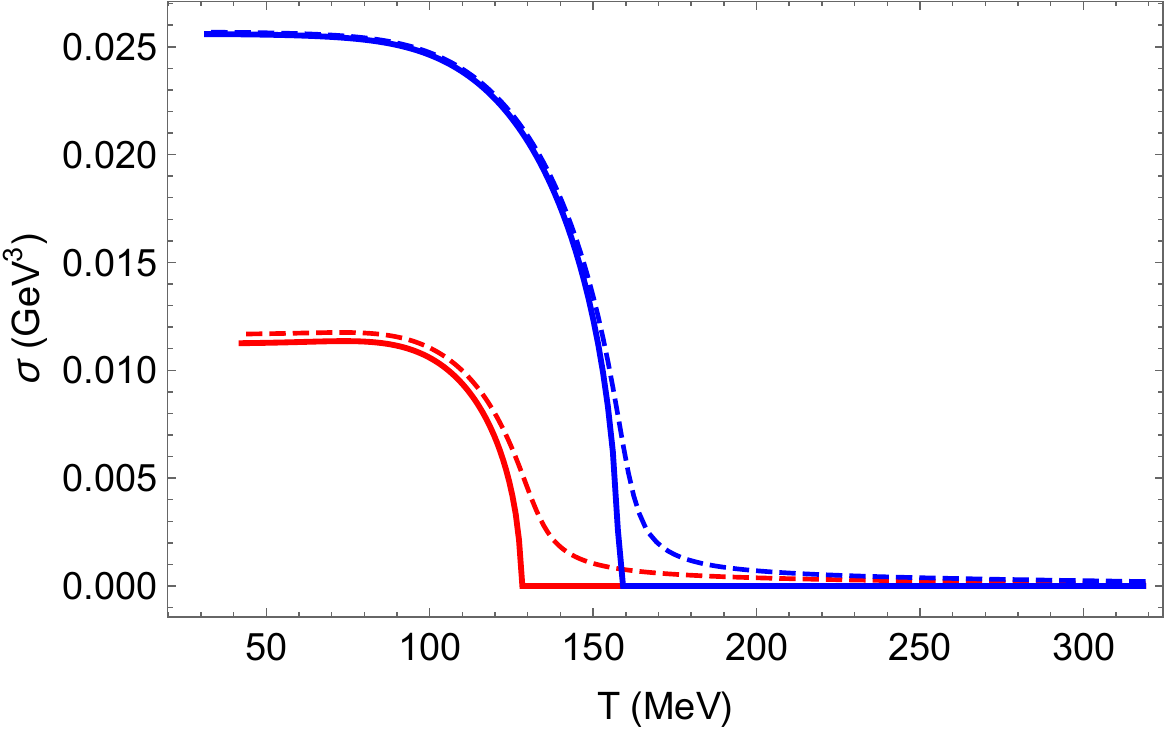}
\hfill
\includegraphics[scale=0.37]{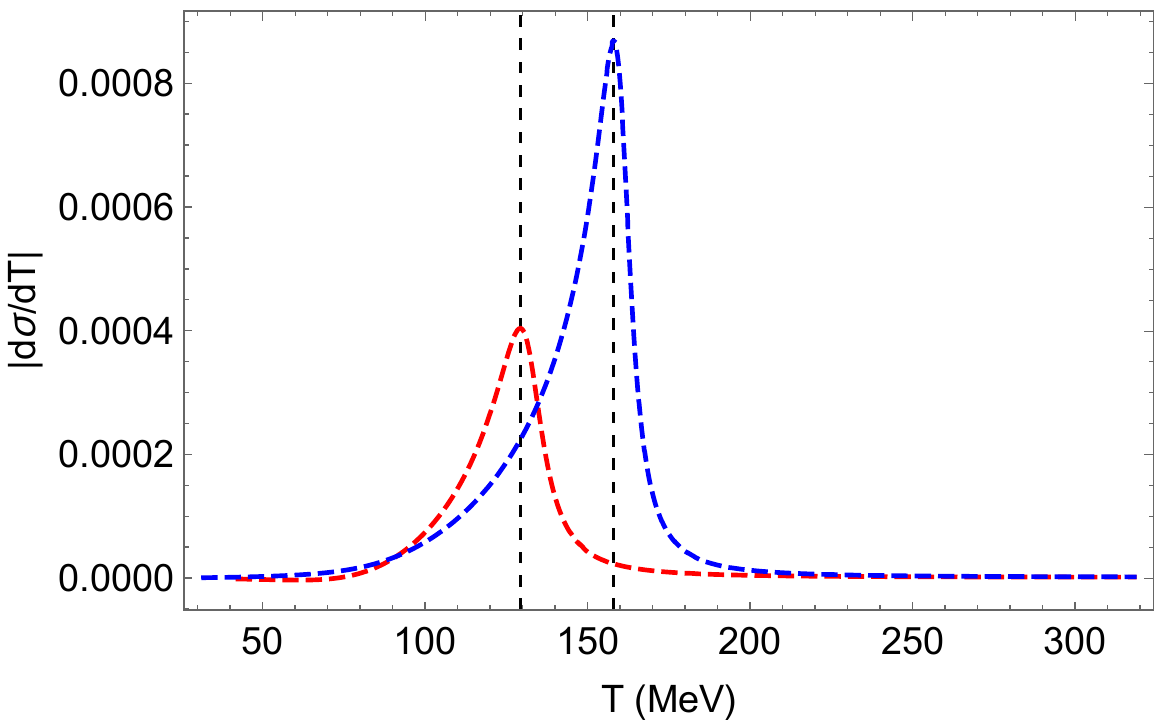}
\includegraphics[scale=0.37]{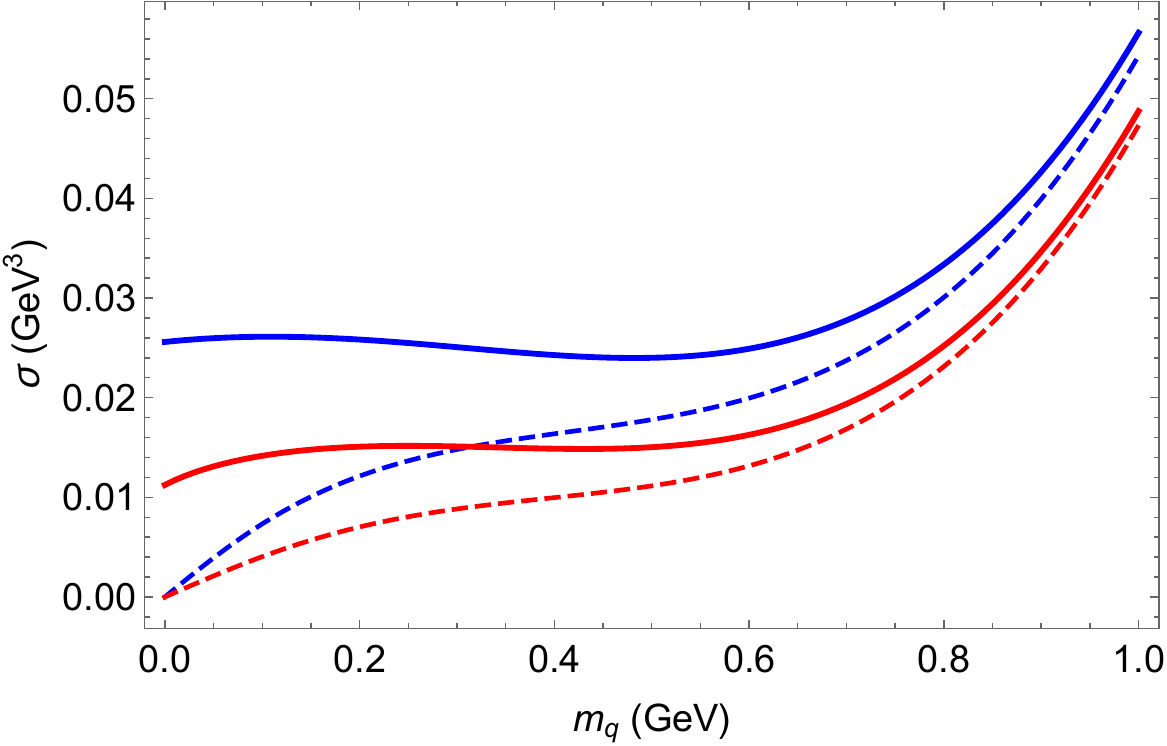}
    \caption{\textbf{Upper Left Panel}: Chiral condensate as a function of the temperature in the chiral limit ($m_{q}=0$).  The blue and red dashed curves represent the chiral condensate as a function of the temperature for finite quark mass, which we fixed as $m_q = 9$ MeV. \textbf{Upper Right Panel}: Absolute value of the derivative of the chiral condensate with respect to the temperature $|\frac{d\sigma}{dT}|$ as a function of the temperature $T$. The dashed vertical lines represent the location of the pseudo-critical temperature $T_{pc}$, which are given by the temperature at the peak of the absolute value of $|\frac{d\sigma}{dT}|$. \textbf{Bottom Panel}: Chiral condensate $\sigma$ as a function of quark mass $m_q$ at $T=0$ (thick red and blue curves) and at $T=200$ MeV (dashed red and blue curves). The red curves correspond to $a_0=6.5$ while the blue ones correspond to $a_0=8.7$.}
    \label{Fig:ChiralTransition}
\end{figure}

In the upper left panel of Fig.~\ref{Fig:ChiralTransition} we display the chiral condensate $\sigma$ as a function of the temperature $T$ in the chiral limit (thick curves) and for finite quark mass (dashed curves).  The red and blue curves correspond to $a_0 = 6.5$ and $a_0 = 8.7$, respectively. One can see that the chiral condensate remains finite for low temperatures, and as temperature increases the chiral condensate vanishes, thus restoring chiral symmetry. Furthermore, the chiral transition in the chiral limit is a second-order phase transition while for finite quark mass is a smooth crossover. This is in accordance with  previous literature \cite{Chelabi:2015cwn,Chelabi:2015gpc} and is consistent with two-flavor QCD \cite{Cuteri:2021ikv}. In the chiral limit, the transition temperature is given by the temperature where the chiral condensate is zero, $\sigma(T_c)=0$. For $a_0 = 6.5$ we find $T_c = 128$ MeV and for $a_0 = 8.7$ we find $T_c = 157$ MeV. We note that the former case is compatible with the recent lattice QCD result in the chiral limit \cite{HotQCD:2019xnw}.
On the other hand, for physical quark mass a pseudo-critical temperature, $T_{pc}$ is obtained from the temperature at the peak of the absolute value of the derivative of the chiral condensate with respect to the temperature, $|\frac{d\sigma}{dT}|$ (shown in the upper right panel of Fig.~\ref{Fig:ChiralTransition}), which is also defined as the susceptibility function. From these peaks we find the pseudo-critical temperature $T_{pc} = 129$ MeV for $a_0 = 6.5$ and $T_{pc} = 158$ MeV for $a_0 = 8.7$. In the latter case,  the corresponding pseudo-critical temperature matches the recent result from lattice QCD for physical quark masses \cite{Borsanyi2020}.

Finally, in the bottom panel of Fig.~\ref{Fig:ChiralTransition} we display the chiral condensate $\sigma$ as a function of quark mass $m_q$ at zero (thick red and blue curves) and finite temperature (such that $T>T_c$), represented by the dashed red and blue curves. The behavior of the chiral condensate $\sigma$ as a function of the quark mass $m_q$ at zero temperature was investigated in our previous work \cite{Ballon-Bayona:2021ibm} and demonstrates spontaneous chiral symmetry breaking in the chiral limit, meaning $\sigma \neq 0$ at $m_q=0$. However, for the chosen temperature in the plot, $T=200$ MeV, which is larger than the critical temperature,  the figure shows that the chiral symmetry has indeed been restored in the chiral limit, that is, $\sigma=0$ at $m_q=0$. This serves as supporting evidence for our numerical results.

\begin{figure}[htb]
    \centering
    \includegraphics[scale=0.37]{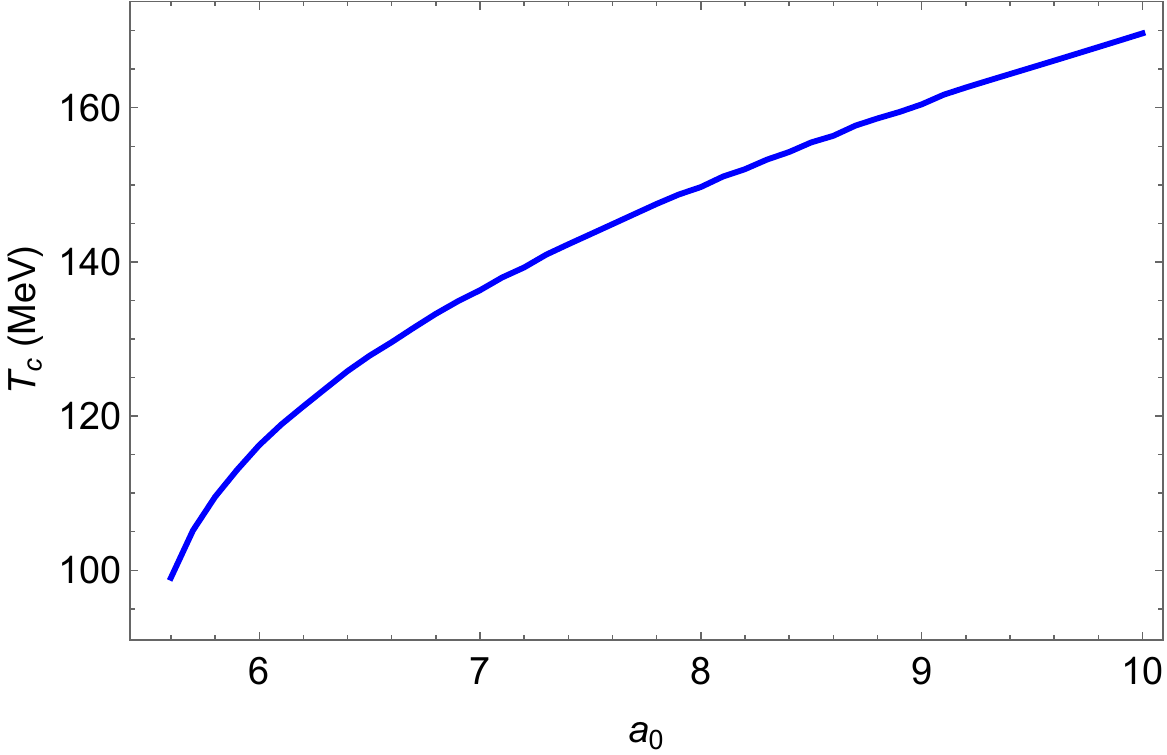}\hfill
    \includegraphics[scale=0.37]{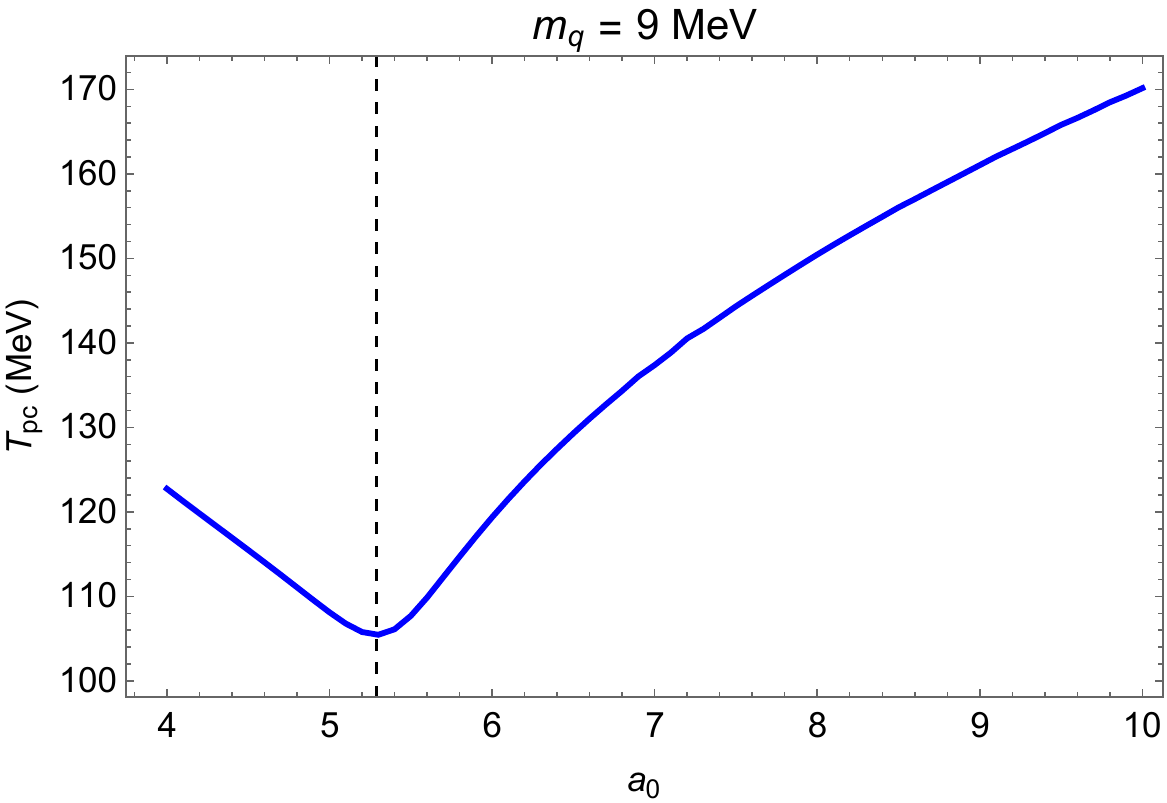}
    \caption{\textbf{Left Panel}: The critical temperature in the chiral limit as a function of $a_0$.  \textbf{Right Panel}: The pseudo-critical temperature at physical quark mass ($m_q = 9$ MeV)  as a function of $a_0$.}
    \label{Fig:Tc_vs_a0}
\end{figure}

In Fig.~\ref{Fig:Tc_vs_a0} we display the behavior of the critical temperature as a function of the dilaton coupling parameter $a_0$ in the chiral limit (left panel) and for finite quark mass (right panel). In the case when $m_q=0$ (chiral limit),  the quark condensate is always zero for $a_0<5.6$ meaning that in this range of the dilaton coupling parameter we are always in the chirally symmetric phase. The corresponding critical temperature $T_c=0$ is not shown in the figure.   On the other hand, for $a_0 > 5.6$,  the critical temperature jumps to $100$ MeV and then increases with $a_0$ as shown in the left panel of Fig.~\ref{Fig:Tc_vs_a0}. In the case of physical quark mass ($m_q = 9$ MeV), there is no restriction on the value of the dilaton coupling parameter $a_0$ and we find that the pseudo-critical temperature displays a non-monotonic dependence on $a_0$, as shown in the right panel of Fig.~\ref{Fig:Tc_vs_a0}. We call the point that separates the two monotonic behaviors the inflection point, $a_{0}^{in} \approx 5.3$. For $a_0<a_{0}^{in}$, the pseudo-critical temperature decreases as we increase $a_0$, while for $a_0>a_{0}^{in}$ the pseudo-critical temperature increases with increasing $a_0$ in the same qualitative way as in the chiral limit.

\subsection{Free energy density}
\noindent

In order to compare the stability of different solutions for the chiral condensate  we need to calculate the
the free energy associated with the tachyonic field in the black brane background. For this purpose, we need to consider the Euclidean on-shell action ($S^{E} = -i\,S$) by  performing an analytical continuation to Euclidean time ($\tau = i\,t$), whose period $\beta$ is the inverse temperature, $\beta = 1/T$. Then, we make the identification $F=T S^{E}$, where $F$ is the free energy.

\begin{figure}[htb]
    \centering
    \includegraphics[scale=0.37]{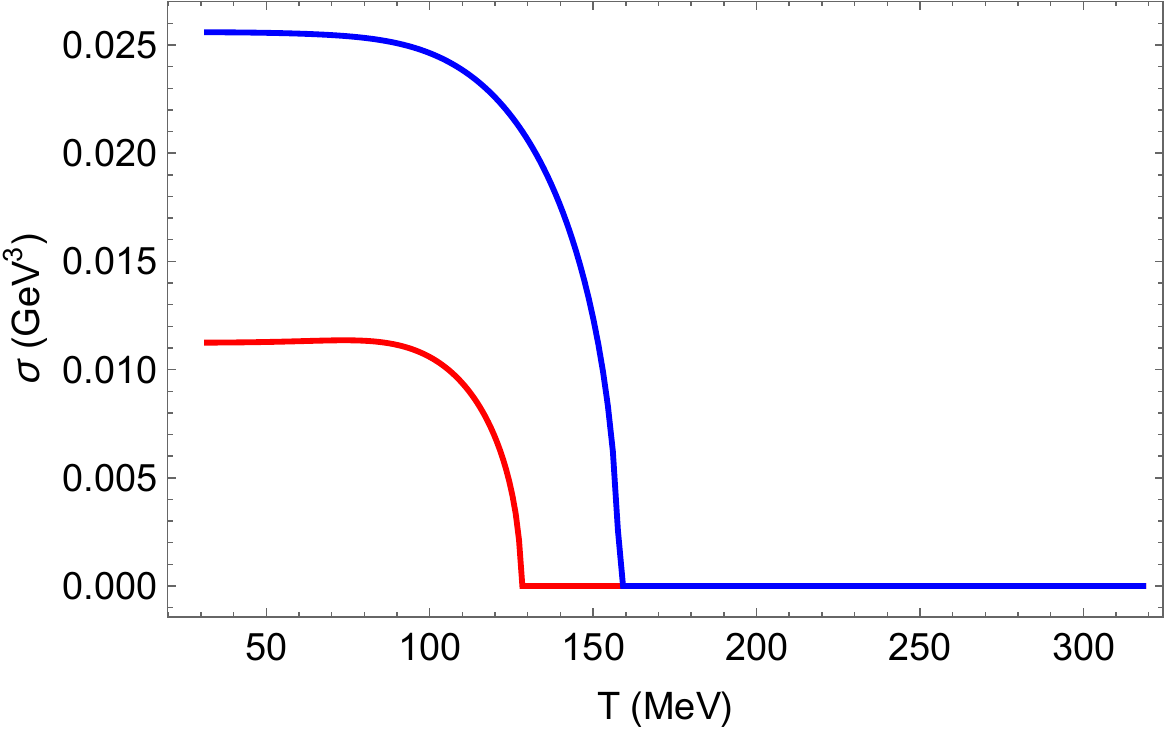}
    \hfill
    \includegraphics[scale=0.37]{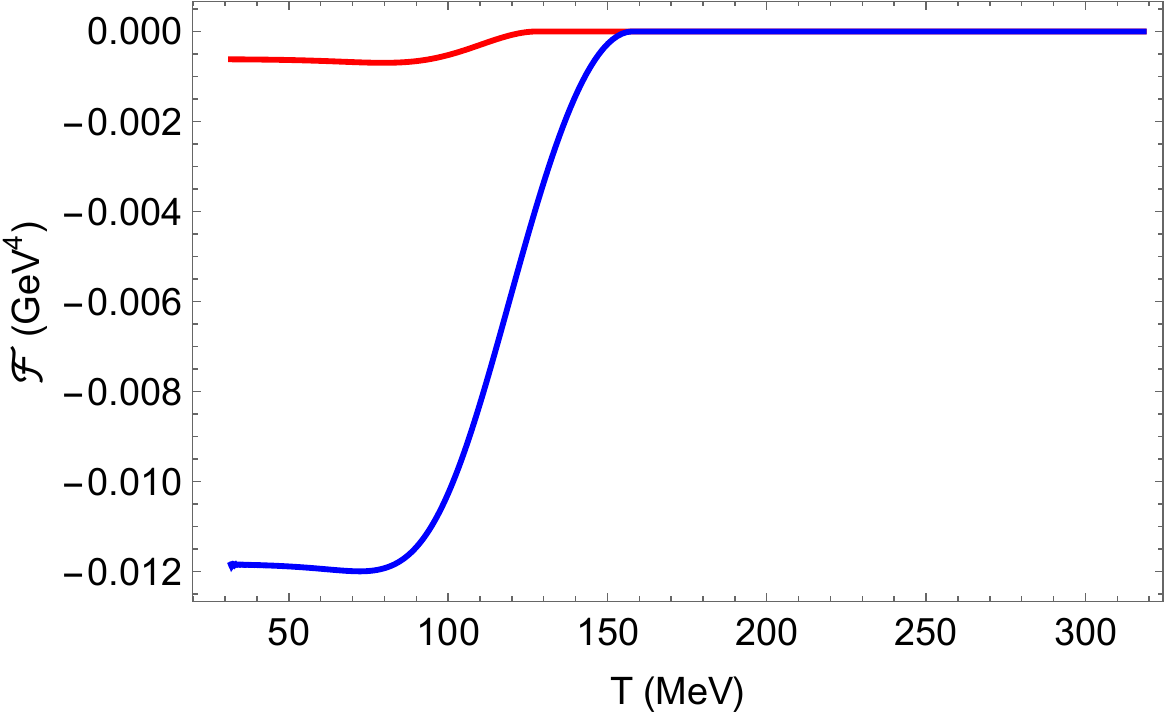}
    \hfill
    \includegraphics[scale=0.37]{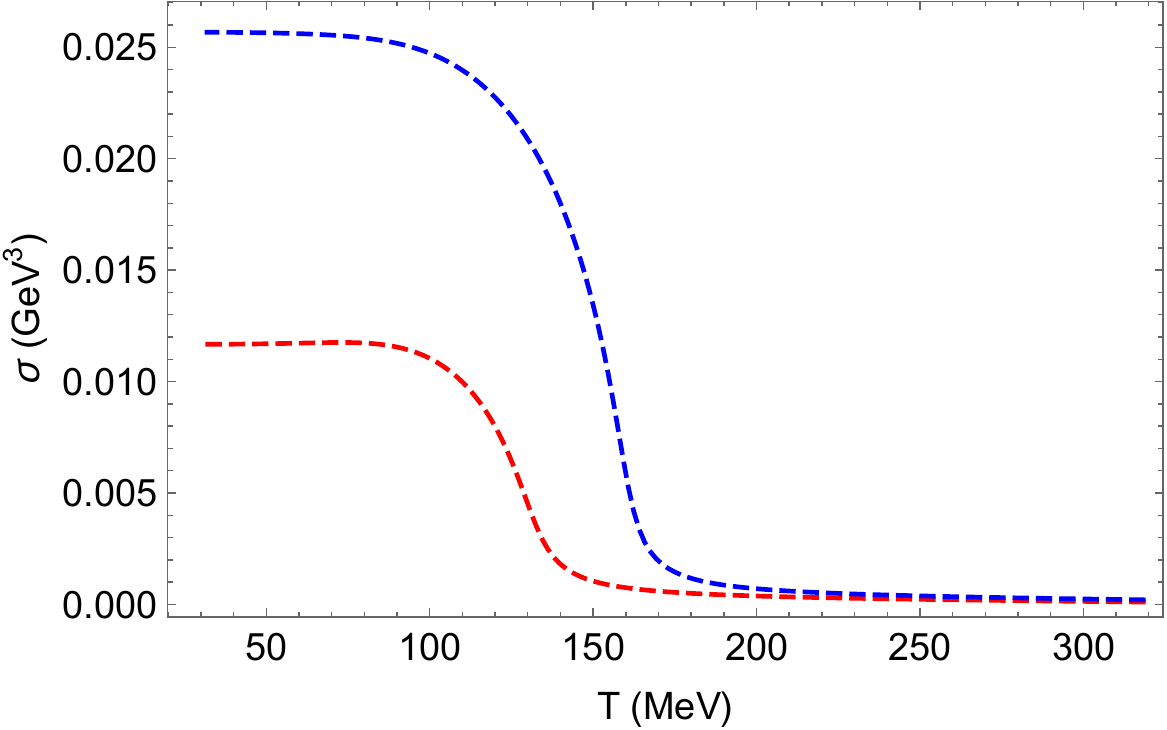}
    \hfill
    \includegraphics[scale=0.37]{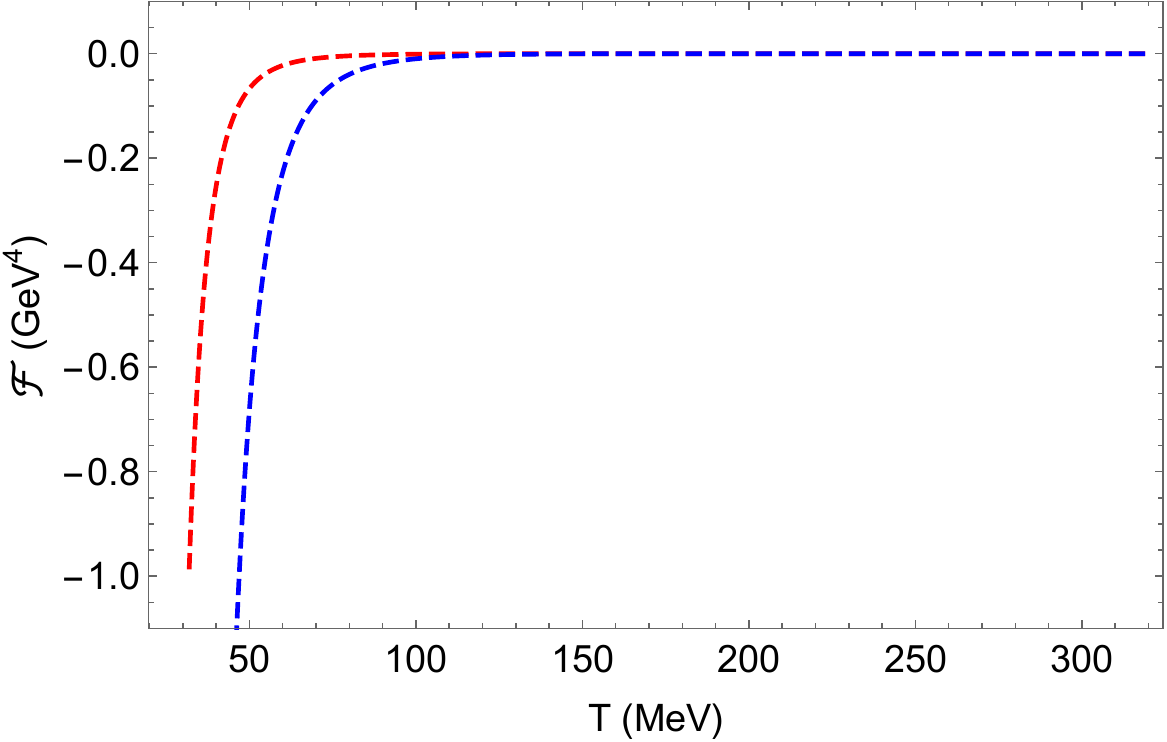}
    \hfill
    \includegraphics[scale=0.37]{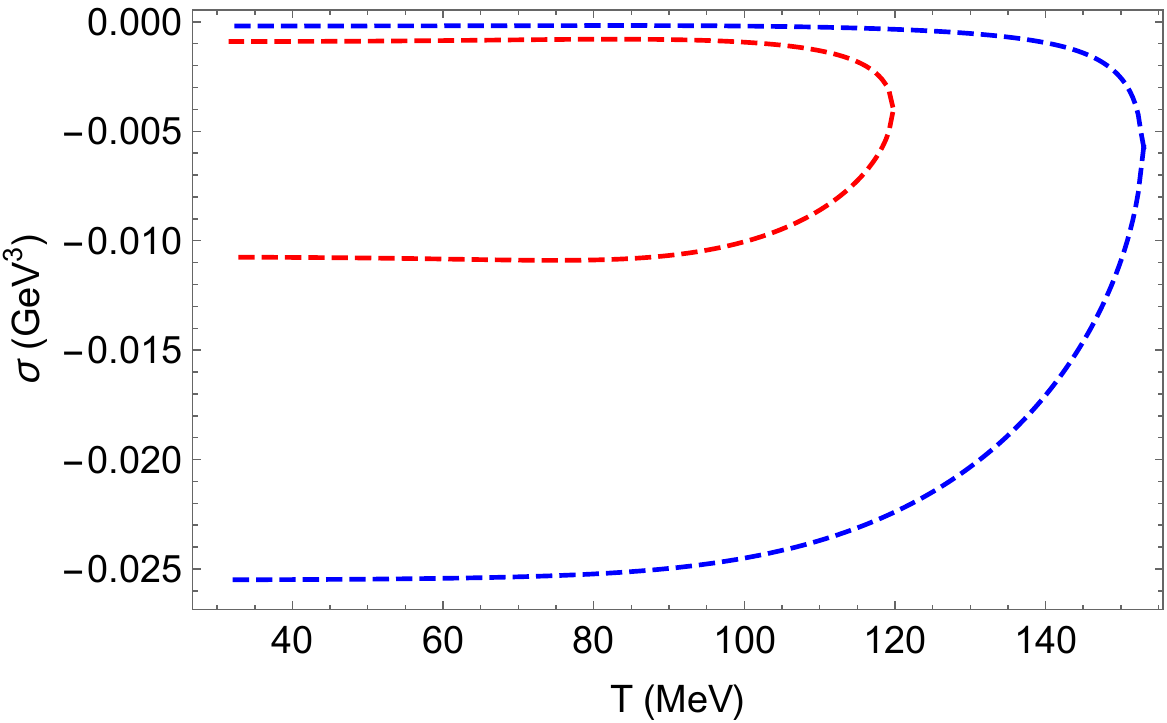}
    \hfill
    \includegraphics[scale=0.37]{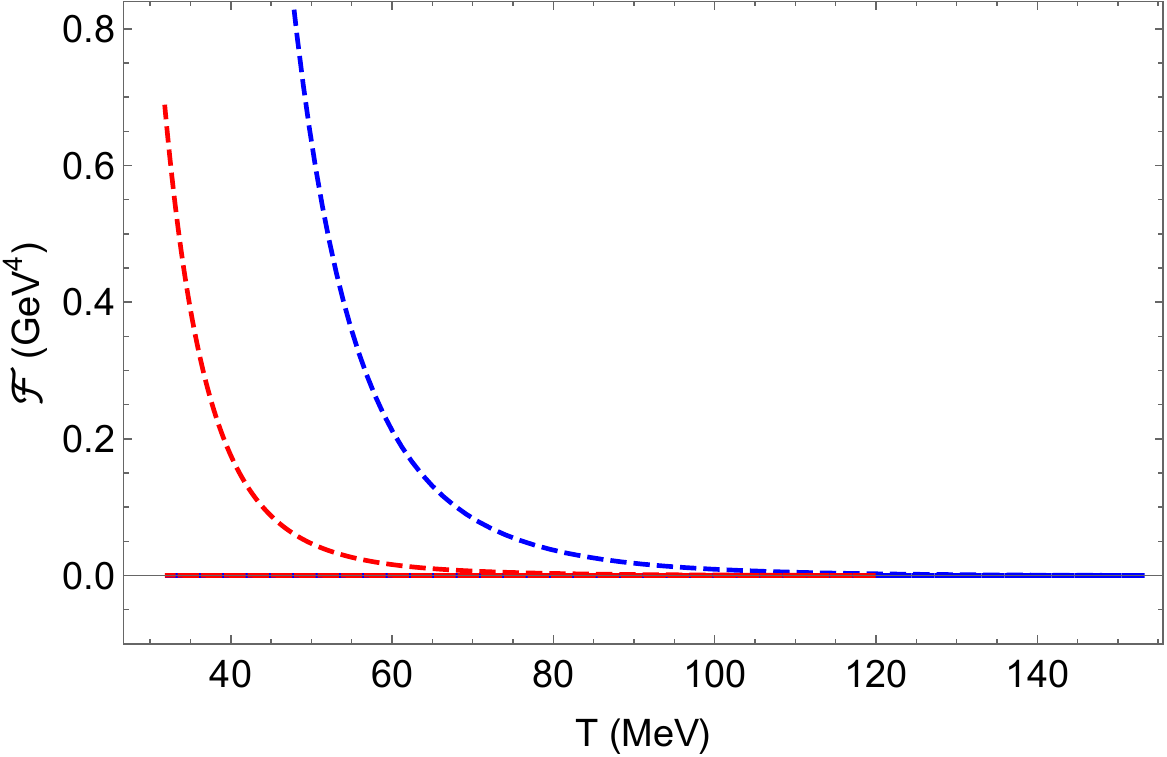}
    
    \caption{\textbf{Upper Panel}: Chiral condensate as a function of the temperature (left) and the free energy as a function of the temperature (right) for $a_0=6.5$ (red curve) and $a_0=8.7$ (blue curve) in the chiral limit. \textbf{Middle Panel}: Chiral condensate as a function of the temperature (left) and the free energy as a function of the temperature (right) for $a_0=6.5$ (red curve) and $a_0=8.7$ (blue curve) for finite quark mass.  \textbf{Lower Panel}: Unphysical solution for the chiral condensate as a function of the temperature (left) and its free energy as a function of the temperature (right) for $a_0=6.5$ (red curve) and $a_0=8.7$ (blue curve) for finite quark mass.}
\label{Fig:Freeenergy}
\end{figure}

The effective Euclidean on-shell action for the tachyon can be written as
\begin{align}
   F = T S^{E} &= -V_{3}\int^{z_h}_{\epsilon} dz \, e^{3A_s - a} \Big [ \frac12 f (\partial_z \chi)^2 + e^{2A_s} V(\chi)  \Big ] \\
    &= -V_{3} \Big [ \dfrac{1}{2}e^{3A_{s}-a}f\chi'\chi\Big|_{\epsilon}+\dfrac{\lambda}{8}\int^{z_h}_{\epsilon} dz \,e^{5A_{s}-a}\chi^4 \Big ],
\end{align}
where $V_3$ is the three-dimensional spatial volume, and $\epsilon$ is a UV cutoff to regularize the Euclidean on-shell action. We take the limit $\epsilon\to0$ at the end of the computation. Note that we  dropped the horizon contribution from the first term in the second line since, by definition, the horizon function $f$ obeys $f(z_h)=0$.\footnote{Since $f(z=\epsilon\to0)=1$, we can also set $f=1$ on the boundary.} By defining the free energy density $\mathcal{F}:=F/V_3$ we have
\begin{equation}\label{freeenergy}
\mathcal{F}  = -\Big [ \dfrac{1}{2}e^{3A_{s}-a}\chi'\chi\Big|_{\epsilon}+\dfrac{\lambda}{8}\int^{z_h}_{\epsilon} dz \,e^{5A_{s}-a}\chi^4 \Big ].
\end{equation}
One can see from the free energy expression above that, in the chiral limit $m_{q}\to 0$, the first term in \eqref{freeenergy} vanishes since $\chi'\propto m_{q}$. Therefore, the solution with $\chi = 0$, which is one trivial vacuum solution in \eqref{tachyoneq}, has zero free energy $\mathcal{F}=0$. This configuration corresponds to the chirally restored phase, with vanishing chiral condensate $\sigma = 0$ at $m_q = 0$. On the other hand, in the spontaneously broken phase, the other trivial vacuum solutions correspond to $\chi = \pm\sqrt{\frac{6}{\lambda}}$ with $\sigma\neq 0$. In this case, for positive quartic coupling ($\lambda >0$) we have from \eqref{freeenergy} that $\mathcal{F}<0$. Therefore, in the chiral limit we expect  the numerical solution for the tachyon with $\sigma\neq0$ to be thermodynamically favored in the vacuum ($T=0$) and in the low temperature regime. This is verified numerically, as shown in the upper right panel of Fig.\ref{Fig:Freeenergy}, which shows the free energy $\mathcal{F}$ as a function of the temperature $T$ in the chiral limit. This is consistent with previous holographic interpolating soft wall models proposed in \cite{Chelabi:2015gpc}. It is worth noting that, as expected for a second-order phase transition, the free energy is an analytic function of the temperature.

For finite quark mass, both the boundary and the interaction terms in \eqref{freeenergy} are divergent. Therefore, one has to add counterterms to cancel these divergences. The specific form of counterterms to tackle the infinities of the on-shell tachyon action were worked out in \cite{Ballon-Bayona:2020qpq}.  By taking the minimal subtraction scheme, we add these counterterms in order to extract the finite part of the on-shell action and obtain the renormalized free energy as a function of the temperature. The results are displayed in the right middle panel of Fig.\ref{Fig:Freeenergy} for physical quark mass $m_q=9$ MeV and $a_0 = 6.5$ (dashed red curve) and $a_0 = 8.7$ (dashed blue curve). As expected, the free energy density is negative for $T< T_{pc}$ signifying that the chirally broken phase is thermodynamically favored at low temperatures.  Since for finite quark mass the chiral transition is a smooth crossover, the free energy is also an analytic function of the temperature.

Finally, for finite quark mass we find another non-trivial solution which appears only for $T<T_{pc}$ and  is characterized by a negative quark condensate. This unphysical solution is shown on the lower left panel of Fig.\,\ref{Fig:Freeenergy}. It is unphysical because its free energy is positive, as shown in the right lower panel of Fig.\,\ref{Fig:Freeenergy}, and therefore it is not thermodynamically preferred compared to the physical solution, which is shown on the right middle panel of Fig.\,\ref{Fig:Freeenergy}. In addition, in the Appendix\,\ref{App:Condensate}, we comment about an issue at finite quark mass which seems to be present in all interpolated softwall models presented so far in the literature, which is the fact that the chiral condensate becomes negative for $T>>T_{pc}$ and increases in absolute value as we approach high temperatures. In the end we make a comparison of the results in our model with the results of \cite{Chelabi:2015cwn,Chelabi:2015gpc} when extrapolated to high temperatures, i.e, for $T>>T_{pc}$. 

\section{Mesons in a deconfined plasma} 
\label{Sec:mesons}

 It is known from QCD that the strong force becomes weaker at high temperatures and energy densities. Under these extreme conditions, hadrons split up into their constituent quarks and gluons, producing the quark-gluon plasma \cite{Hwa:2010npa}. The melting of mesons is a key process to understand the physical properties of the quark-gluon plasma. This problem has been investigated using the gauge/gravity duality, see for instance \cite{Hoyos-Badajoz:2006dzi,Peeters:2006iu,Fujita:2009ca,Miranda:2009uw, Mamani:2013ssa,Cui:2014oba,Grigoryan:2010pj,Dudal:2014jfa,Bartz:2016ufc,Braga:2017bml,Mamani:2018uxf}. 
 
 In this section we describe the dynamics of the 5d field perturbations associated with 4d vector and scalar mesons. We will investigate the effective Schr\"odinger potentials for the vectorial and scalar sector and conclude from that analysis that the mesons melt at sufficiently large temperatures.  We will also perform a hydrodynamic expansion in the longitudinal part of the vectorial sector that will allow us to extract the diffusion constant for a flavor current in the plasma. The melting temperatures for the scalar and vector mesons will be  obtained in the next section from an analysis of the meson spectral functions. 

\subsection{Vector mesons} \label{sec:vector_melting}

We begin with the portion of the action \eqref{Eq:5dmodel} that describes the gauge field perturbations $V_m = ( \delta A_m^{(R)} + \delta A_m^{(L)})/2$ associated with the vector mesons, 
\noindent
\begin{equation}
    S_V=-\int d^5x\sqrt{-g} \frac{e^{-b(\Phi)}}{4g_5^2}F_{mn}F^{mn}
\end{equation}
\noindent
where $F_{mn}=\partial_mV_n-\partial_nV_m$, and $V_m$ is the vector potential. As described in section \ref{Sec:Model}, we choose to couple the dilaton to the scalar and vector sectors in the same way, that is, $b(\Phi)=a(\Phi)$ given in \eqref{ModelIIA}. Then, the equations of motion are given by
\noindent
\begin{equation}
    \partial_m\left(\sqrt{-g} \, e^{-a(\Phi)}F^{mn}\right)=0
\end{equation}
\noindent

Considering a gauge field perturbation propagating along the $x^3$ direction we have two sectors: transverse and longitudinal to the direction of propagation. In order to simplify the analysis we consider the Fourier transformation with the wavenumber $k_{\mu}=(-\omega,0,0,k)$. Thus, the equations of motion become
\noindent
\begin{align}
\partial_z^2V_{\alpha}+\left(A_s'+\frac{f'}{f}-a'\right)\partial_zV_{\alpha}+\frac{(\omega^2-k^2f)}{f^2}V_{\alpha}=\,&0,\qquad (\alpha=x^1,x^2) \label{Eq:TransVectorN1}\\
\partial_z^2V_{x^0}+\left(A_s'-a'\right)\partial_zV_{x^0}-\frac{k^2}{f}V_{x^0}-\frac{k\,\omega}{f}V_{x^3}=\,&0, \label{Eq:LongVectorN1}\\
\partial_z^2V_{x^3}+\left(A_s'+\frac{f'}{f}-a'\right)\partial_zV_{x^3}+\frac{k\,\omega}{f^2}V_{x^0}+\frac{\omega^2}{f^2}V_{x^3}=\,&0, \label{Eq:LongVectorN2} \\
kf\partial_zV_{x^3}+\omega \partial_zV_{x^0}=\,&0. \label{Eq:LongVectorN3}
\end{align}
\noindent
The transverse sector is described by the differential equation \eqref{Eq:TransVectorN1}, while the longitudinal sector is described by the coupled differential equations \eqref{Eq:LongVectorN1}, \eqref{Eq:LongVectorN2} and \eqref{Eq:LongVectorN3}. 

It is useful to write these equations in terms of gauge invariant fields \cite{Kovtun:2005ev, Mamani:2013ssa}. For the transverse and longitudinal sectors we define the gauge invariant fields:
\noindent
\begin{equation}\label{Eq:GaugeInvariantEqs}
    E_\alpha=\omega\,V_{\alpha},\qquad\qquad E_{x^3}=\omega\,V_{x^3}+k\,V_{x^0}.
\end{equation}
\noindent
Thus, for the transverse sector the differential equation in terms of the gauge invariant field becomes 
\noindent
\begin{align}
    \partial_z^2E_{\alpha}+\left(A_s'+\frac{f'}{f}-a'\right)\partial_zE_{\alpha}+\frac{(\omega^2-k^2f)}{f^2}E_{\alpha}=\,&0.\qquad (\alpha=x^1,x^2) \label{Eq:TransVectorN2}
\end{align}
\noindent
In the longitudinal sector, the coupled equations \eqref{Eq:LongVectorN1}, \eqref{Eq:LongVectorN2} and \eqref{Eq:LongVectorN3} are reduced to a single  equation for the gauge invariant field $E_{x^3}$, given by
\noindent
\begin{equation}\label{Eq:LongGaugeVector}
    \partial^2_zE_{x^3}+\left(A_s'-a'+\frac{\omega^2 f'}{f \left( \omega^2-k^2f \right)}\right)\partial_zE_{x^3}+\frac{\left(\omega^2-k^2f\right)}{f^2}E_{x^3}=0.
\end{equation}
\noindent
When the spatial momentum $k=0$, Eq.~\eqref{Eq:LongGaugeVector} reduces to  Eq.~\eqref{Eq:TransVectorN2}, and the transverse and longitudinal sectors have the same behavior, as expected. 

In order to solve these differential equations it is useful to write them in the so-called Schr\"odinger-like form \cite{Miranda:2009uw, Mamani:2013ssa, Mamani:2022qnf}. We use the tortoise coordinate \eqref{torteq}, and switch to the dimensionless coordinate $u=\phi^{1/2}_\infty\,z$, frequency $\overline{\omega}=\phi_{\infty}^{-1/2}\omega$ and wavenumber $\overline{k}=\phi_{\infty}^{-1/2}k$. 

Using the Bogoliubov transformations $E_{\alpha}=e^{-B_T}\psi_T$ and $E_{x^3}=e^{-B_L}\psi_L$, with $2B_T=A_s-a$ and $2B_L=A_s-a-\ln{\left(\overline{\omega}^2-  \overline{k}^2f\right)}$, the differential equations \eqref{Eq:TransVectorN2} and \eqref{Eq:LongGaugeVector} take on the Schr\"odinger-like form
\noindent
\begin{align}
-\partial_{\, \overline{r}_*}^{2}\psi_{j}+V_{T/L}\,\psi_{j}=\overline{\omega}^2\psi_{j}  \quad\qquad (j=x^1,x^2,x^3)\label{eq:SchrodingerVector}
\end{align}
\noindent
where $\, \overline{r}_* = \phi^{1/2}_\infty\, r_*$ and $V_T$ ($V_L$) is the transverse (longitudinal) potential given by
\noindent
\begin{equation}\label{Eq:TransvPotential}
V_{T/L}=f(u) \Big \{ \overline{k}^2+(\partial_uf)(\partial_uB_{T/L})+f\left[\left(\partial_uB_{T/L}\right)^2+\partial^2_uB_{T/L}\right] \Big \}.
\end{equation}
\noindent
This form makes it clear that the potential is zero at the horizon, as $f(u_h)=0$. Thus, \eqref{eq:SchrodingerVector} has the same mathematical structure as the harmonic oscillator problem, so the near-horizon solutions are
\noindent
\begin{equation}
    \psi_j\sim \mathfrak{C}_je^{-i\overline{\omega} \, \overline{r}_*}+\mathfrak{D}_je^{i\overline{\omega} \, \overline{r}_*}. \label{eq:incomingOutgoing}
\end{equation}
\noindent
The negative exponential is interpreted as an incoming wave at the horizon, while the positive exponent is interpreted as an outgoing wave coming from the black hole interior, which is forbidden from the classical point of view. From the point of view of the gauge/gravity duality, the incoming solution is related to retarded correlation functions, while the outgoing is related to advanced correlation functions \cite{Son2002,Kovtun:2005ev}. 

In order to calculate the spectral functions, we need the asymptotic solutions close to the horizon, which are written as 
\noindent
\begin{equation}
    \psi_j^{\pm}=e^{\pm i\,\overline{\omega} \, \overline{r}_*}\left(a_{j,0}^{(\pm)}+a_{j,1}^{(\pm)}\left(u_h-u\right)+a_{j,2}^{(\pm)}\left(u_h-u\right)^2+\cdots\right),
\end{equation}
where the coefficients are 
\noindent
\begin{equation}
    \begin{split}
         a_{j,1}^{(\pm)}=\,&\frac{a_{j,0}^{(\pm)}u_h^2}{2(2\pm\overline{\omega} u_h i)} \Big [ \frac{2}{u_h^3}+\frac{4+\overline{k}^2}{u_h}+\frac{8a_0u_h^4}{\left(1+u_h^4\right)^2}-\frac{6a_0}{1+u_h^4}+\frac{8\overline{k}^2}{u_h^3\overline{\omega}^2}\delta_{jx^3} \Big ] \, ,
    \end{split}
\end{equation}
\noindent
\noindent
\begin{equation}
    \begin{split}
        a_{j,2}^{(\pm)}=\,&\frac{u_h^2}{16(4\pm\overline{\omega} u_h i)}\Big \{ 2a_{j,0}^{(\pm)}\bigg[8-\frac{8}{u_h^4}-\frac{44+3\overline{k}^2}{u_h^2}+\frac{32a_0^2u_h^{10}}{\left(1+u_h^4\right)^4} +\frac{12a_0}{u_h(1+u_h^4)}\\
        \,&-\frac{(266+48a_0u_h^3+10u_h^4)a_0u_h^3}{\left(1+u_h^4\right)^3}+\frac{(78+18a_0u_h^3+32u_h^6)a_0}{u_h\left(1+u_h^4\right)^2}-\frac{24a_0u_h}{1+u_h^4}\\
        &+\left(\frac{32\overline{k}^2}{u_h^2\overline{\omega}^2}+\frac{64a_0u_h^3\overline{k}^2}{(1+u_h^4)^2\overline{\omega}^2}+\frac{160\overline{k}^2}{u_h^4\overline{\omega}^2}-\frac{152\overline{k}^2}{u_h^2\overline{\omega}^2}-\frac{48a_0\overline{k}^2}{u_h(1+u_h^4)\overline{\omega}^2}\right)\delta_{jx^3}\bigg]\\
        \,&-4a_{j,1}^{(\pm)}\bigg[\frac{20}{u_h^3}+\frac{4+\overline{k}^2}{u_h}+\frac{8a_0u_h^4}{(1+u_h^4)^2}-\frac{6a_0}{1+u_h^4}\pm\frac{3i\overline{\omega}}{u_h^2}+\frac{8\overline{k}^2}{\overline{\omega}^2u_h^3}\delta_{jx^3}\bigg] \Big \} \, ,
    \end{split}
\end{equation}
\noindent
where $\delta_{jx^3}$ is the Kronecker delta. 

Meanwhile, the asymptotic solutions close to the boundary are given by
\noindent
\begin{equation}
\label{eq:Vector_asymptotic_boundary}
    \begin{split}
        \psi^{(1)}_j=\,&u^{3/2}\left(b_{j,0}+b_{j,1}u+b_{j,2}u^2+b_{j,3}u^3+b_{j,4}u^4+\cdots\right)\\
        \psi^{(2)}_j=\,&u^{-1/2}\left(c_{j,0}+c_{j,1}u+c_{j,2}u^2+c_{j,3}u^3+c_{j,4}u^4+\cdots\right)+2d_j\psi^{(1)}_j\ln{[u]},
    \end{split}
\end{equation}
where $\psi^{(1)}_j(u)$ is the normalizable solution, while $\psi^{(2)}_j(u)$ is the non-normalizable. The general solution can be written as $\psi_j = \psi^{(1)}_j + \psi^{(2)}_j$  and should contain only two independent coefficients, namely $b_{j,0}$ and $c_{j,0}$. The coefficient $c_{j,2}$ is redundant because all the terms arising from it are identical to the terms arising from $b_{j,0}$ and we therefore set $c_{j,2}$ to zero. The other coefficients are given by: 
\begin{equation} \label{eq:Vector_coeff}
    \begin{split}
        b_{j,1}=\,&0,\qquad b_{j,2}=\frac{b_{j,0}}{8}\left(\overline{k}^2-\overline{\omega}^2\right),\qquad b_{j,3}=\frac{a_{0}\,b_{j,0}}{10}\\
        b_{j,4}=\,&\frac{b_{j,0}}{192}\left(8+\frac{64}{u_h^4}+\left(\overline{k}^2-\overline{\omega}^2\right)^2+\frac{32\overline{k}^2}{u_h^4(\overline{k}^2-\overline{\omega}^2)}\delta_{jx^3}\right)\\
        {c}_{j,1}=\,& 0,\qquad c_{j,3}=\frac{c_{j,0}}{2}a_0,\\
        c_{j,4}=\,&\frac{c_{j,0}}{16}\left(2-\frac{3}{4} \left(\overline{k}^2-\overline{\omega}^2\right)^2+\frac{8\overline{k}^2}{u_h^4(\overline{k}^2-\overline{\omega}^2)}\delta_{jx^3}\right)\\
        d_j=\,& \frac{c_{j,0}}{4b_{j,0}}\left(\overline{k}^2-\overline{\omega}^2\right)
    \end{split}
\end{equation}
The asymptotic solutions \eqref{eq:incomingOutgoing} and \eqref{eq:Vector_coeff} will be used in section \ref{sec:spectral_functions} to calculate the vector spectral functions. 

\subsection{The hydrodynamic diffusion constant of a flavor current}

Using the gauge/gravity duality one can get additional physical properties of the deconfined plasma, such as the computation of some transport coefficients in the hydrodynamic regime. In this regime the frequency and wavenumber are smaller than the temperature, i.e., $\omega\ll T$ and $k\ll T$. The method for calculating hydrodynamic transport coefficients from the dynamics of field perturbations in the black brane was proposed in the seminal papers \cite{Policastro:2002tn,Policastro:2002se};  for applications see for example \cite{Mamani:2013ssa, Mamani:2018qzl, Diles:2019uft, Mamani:2022qnf}. There is a complementary approach to extract the transport coefficients which involves the calculation through Kubo formulas, see for instance \cite{Li:2014dsa, Finazzo:2014cna, Gubser:2008sz, Gursoy:2009kk, Finazzo:2016mhm,Ballon-Bayona:2021tzw} and references therein.

In order to find the diffusion constant we solve the differential equation \eqref{Eq:LongGaugeVector} in the hydrodynamic limit. First, we define dimensionless ratios for the frequency $\omega$ and the wave number $k$ dividing by the temperature as $\mathfrak{w}=\frac{\omega}{\pi T}$, $\mathfrak{q}=\frac{k}{\pi T}$. We also a define dimensionless ratio for the dilaton mass constant $\mathfrak{c}=\frac{\phi_{\infty}}{\pi^2T^2}$ and for the radial coordinate $\mathfrak{u}=z\,\pi T$. Note that the definition of the temperature \eqref{Eq:Temperature} allows us to write $z_h=\frac{1}{\pi T}$, which means that the horizon is located at $\mathfrak{u}_h=1$. Considering the incoming solution at the horizon, we can write the gauge field in terms of a regular function, $F(\mathfrak{u})$, at the horizon in the form
\noindent
\begin{equation}
    E_{x^3}=f^{-\frac{i \mathfrak{w}}{4}}F_{x^3}(\mathfrak{u}),
\end{equation}
\noindent
In order to simplify the analysis we introduce a new rescaling of the relevant parameters: $\mathfrak{w}\to \lambda\mathfrak{w}$ and $\mathfrak{q}\to \lambda\mathfrak{q}$ with $\lambda \ll 1$, with $\lambda$ the relevant parameter in the forthcoming analysis \cite{Kovtun:2005ev, Mamani:2022qnf, Mamani:2018qzl}. Thus, the differential equation for $F_{x^3}(\mathfrak{u})$ is
\noindent
\begin{equation}
    \begin{split}
        &F''_{x^3}+\Big[\frac{32\mathfrak{q}^2\mathfrak{w}^2f-16\mathfrak{w}^4-16\mathfrak{q}^4f^2+4i\mathfrak{q}^3fA_s'f'-4i\mathfrak{q}^2\mathfrak{w}f^2A_s'f'-4i\mathfrak{w}^3fa'f'}{16f^2(\mathfrak{q}^2f-\mathfrak{w}^2)}\\
        &+\frac{4i\mathfrak{q}^2\mathfrak{w}f^2a'f'+\mathfrak{w}^4f'^{\,2}+4i\mathfrak{q}^2\mathfrak{w}ff'^{\,2}-\mathfrak{q}^2\mathfrak{w}^2ff'^{\,2}}{16f^2(\mathfrak{q}^2f-\mathfrak{w}^2)}-\frac{i\mathfrak{w}f''}{4f}\Big]F'_{x^3}\\
        &+\Big[A_s'-a'+\frac{i\mathfrak{w}(\mathfrak{w}(2i+\lambda\mathfrak{w})-\mathfrak{q}^2\lambda f)f'}{2f(\mathfrak{q}^2f-\mathfrak{w}^2)}\Big]F_{x^3}=0.
    \end{split}
\end{equation}
\noindent
Next, we consider the expansion in $\lambda$, $F_{x^3}=F_{x^3,0}(\mathfrak{u})+\lambda F_{x^3,1}(\mathfrak{u})+\lambda^{2} F_{x^3,2}(\mathfrak{u})+\cdots$. By plugging this expansion in the last equation and solving order-by-order in $\lambda$ one gets differential equations for $F_{x^3,0}, F_{x^3,1}, \cdots$. In order to extract the diffusion constant we need the solutions for $F_{x^3,0}$ and $F_{x^3,1}$. Thus, the differential equations we need to solve are:
\begin{align}
    &F_{x^3,0}''+\left[A_s'-a'-\frac{\mathfrak{w}^2f'}{f(\mathfrak{q}^2f-\mathfrak{w}^2)}\right]F_{x^3,0}'=0\label{Eq:F0Long}\\
    &F_{x^3,1}''+\Big[A_s'-a'-\frac{\mathfrak{w}^2f'}{f(\mathfrak{q}^2f-\mathfrak{w}^2)}\Big]F_{x^3,1}'-\frac{i\mathfrak{w}f'}{2f}F_{x^3,0}'+\nonumber\\
    &\Big[\frac{4i\mathfrak{w}f\left(\mathfrak{w}^2A_s'f'-\mathfrak{w}^2a'f'+\mathfrak{q}^2f'^{\,2}+\mathfrak{w}^2f''\right)-4i\mathfrak{q}^2f^2\left(\mathfrak{w}A_s'f'-\mathfrak{w}a'f'+\mathfrak{w}f''\right)}{16f^2(\mathfrak{q}^2f-\mathfrak{w}^2)}\Bigg]F_{x^3,0}=0.\label{Eq:F1Long}
\end{align}
\noindent
The regular solution of \eqref{Eq:F0Long} is a constant, i.e., $F_{x^3,0}(\mathfrak{u})=F_{x^3,0}$. In turn, the general solution of \eqref{Eq:F1Long} is
\noindent
\begin{equation}
    F_{x^3,1}=C_2+\frac{iF_{x^3,0}\mathfrak{w}}{4}\ln{f}-C_1\mathfrak{q}^2\int_{0}^{\mathfrak{u}}e^{-A_s+a}dx+C_1\mathfrak{w}^2\int_{0}^{\mathfrak{u}}\frac{e^{-A_s+a}}{f}dx,
\end{equation}
\noindent
where $C_1$ and $C_2$ must be fixed by the regularity condition of $F_{x^3,1}$ at the horizon, $\mathfrak{u}_h=1$. Expanding close to the horizon, we get the value for $C_1$ and $C_2$
\noindent
\begin{align}
    C_1=\,&-i\frac{F_{x^3,0}f'(\mathfrak{u}_h)e^{A_s(\mathfrak{u}_h)-a(\mathfrak{u}_h)}}{4\mathfrak{w}}=-i\frac{F_{x^3,0}}{4\mathfrak{w}}\lim_{y\to\mathfrak{u}_h}\frac{\ln{f(y)}}{\int_{0}^{y}\frac{e^{-A_s(x)+a(x)}}{f(x)}dx}\\
    C_2=\,&C_1\mathfrak{q}^2\int_{0}^{\mathfrak{u}_h}e^{-A_s(x)+a(x)}dx
\end{align}
\noindent
Then, the complete solution up to first-order in $\lambda$ is
\begin{equation}
    E_{x^3}=f^{-\frac{i \mathfrak{w}}{4}}\left(F_{x^3,0}+\lambda F_{x^3,1}(\mathfrak{u})\right)
\end{equation}
\noindent
Expanding close to the boundary and imposing Dirichlet boundary conditions we obtain the dispersion relation (setting the auxiliary parameter $\lambda\to1$ and restoring the original parameters)
\noindent
\begin{equation}
    \omega=\frac{if'(\mathfrak{u}_h)}{4\pi T\mathfrak{u}_h}e^{-a(\mathfrak{u}_h)}\int_{0}^{\mathfrak{u}_h}xe^{a(x)}dx\,k^2.
\end{equation}
\noindent
Now one can read off the diffusion coefficient from the last equation by comparing with Fick's law
\begin{equation}
    \omega=-iDk^2.
\end{equation}
\noindent
Thus, we get
\noindent
\begin{equation}
    D=-\frac{f'(\mathfrak{u}_h)}{4\pi T}e^{-a(\mathfrak{u}_h)}\int_{0}^{\mathfrak{u}_h}x \, e^{a(x)}dx
\end{equation}
\noindent
This result extends the previous result obtained in the soft wall model for light quarks \cite{Mamani:2013ssa}.

\begin{figure}[ht!]
    \centering
    \includegraphics[width=0.6\textwidth]{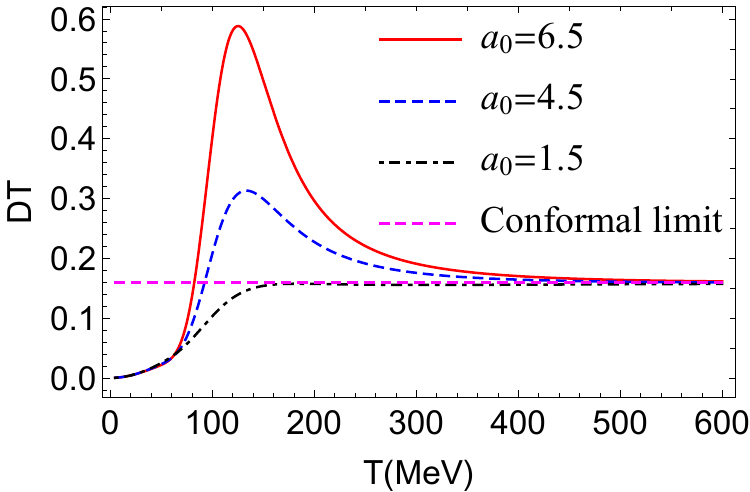}
    \caption{The dimensionless product $DT$, where $D$ is the diffusion constant, as a function of the temperature for different values of the dilaton coupling parameter $a_0$.}
    \label{Fig:DT}
\end{figure}

A plot of the diffusion constant multiplied by the temperature as a function of the temperature is displayed in Fig.~\ref{Fig:DT}, where solid red line represents the result for $a_0=6.5$, dashed blue line for $a_0=4.5$, dashed dotted black line for $a_0=1.5$, and dashed magenta line represents the conformal field theory result, $DT=1/2\pi$. As can be seen, for $a_0=6.5$, the diffusion constant increases with the temperature in the low temperature regime, peaking around $T=125\,\text{MeV}$, then decreasing asymptotically to the conformal value at high temperatures. The temperature where the product $DT$ is a maximum depends on the dilaton coupling parameter, $a_0$, decreasing with the increasing of the dilaton coupling parameter. 
\noindent

%====================================================================
\subsection{Scalar mesons}
\label{SubSec:scalarmesons}

Now we focus on the part of the action \eqref{Eq:5dmodel} that describes the fluctuations in the scalar sector $\delta X = S$, 
\noindent
\begin{equation}\label{Eq:ScalarMesonsEffAct}
    S_S=-2\int dx^5\sqrt{-g}e^{-a(\Phi)}\left[\left(\partial^mS\right)\left(\partial_mS\right)+\left(m_X^2+\frac{3}{2}\lambda \chi^2(z)\right)S^2\right].
\end{equation}
\noindent
The equation of motion is given by 
\noindent
\begin{equation}
    \frac{e^{a}}{\sqrt{-g}}\partial_m\left(\sqrt{-g}e^{-a}g^{mn}\partial_nS\right)-\left(m_X^2+\frac32\lambda\,\chi^2\right)S=0.
\end{equation}
\noindent
As before, considering the direction of propagation along the $x^3$ direction, and the Fourier transformation for the scalar field, the differential equation of the scalar mesons at finite temperature is given by
\noindent
\begin{equation}\label{Eq:ScalarSectorEq}
\partial_u^2S+\left(3A_s'+\frac{f'}{f}-a'\right)\partial_uS+\left(\frac{e^{2A_s}}{f}\left[3-\frac{3}{2}\lambda \chi^2\right]+\frac{\overline{\omega}^2-\overline{k}^2f}{f^2}\right)S=0,
\end{equation}
\noindent
where we use the dimensionless coordinate $u=z\,\phi_{\infty}^{1/2}$, frequency and wavenumber. Using the Bogoliubov transformation $S=e^{-B_S}\psi_S$, where $2B_S=3A_s-a$, we rewrite this equation in the Schr\"odinger-like form
\noindent
\begin{equation}\label{Eq:ScalarSectorSchrodEq}
-\partial_{\, \overline{r}_*}^{2}\psi_S+V_S\,\psi_S=\overline{\omega}^2\psi_S
\end{equation}
\noindent
where the potential, in terms of the tortoise coordinate, is given by
\noindent
\begin{equation}
V_S=\overline{k}^2\,f+e^{2A_s}f\left(-3+\frac{3}{2}\lambda\,\chi^2\right)+\left(\partial_{\, \overline{r}_*}B_S\right)^2+\partial^2_{\, \overline{r}_*}B_S. \label{Eq:ScalarPotential_tortoise}
\end{equation}
\noindent
In terms of the dimensionless coordinate the potential becomes
\noindent
\begin{equation}\label{Eq:ScalarPotential_holographic}
V_S=f\left(\overline{k}^2+e^{2A_s}\left(-3+\frac{3}{2}\lambda\,\chi^2\right)+f\left(\partial_{u}B_S\right)^2+\partial_u\left(f\partial_{u}B_S\right)\right).
\end{equation}
\noindent
Note that in this form, it is easy to see that the potential is zero at the horizon, because $f(u_h)=0$. Thus, the Schr\"odinger-like equation has the solutions
\noindent
\begin{equation}\label{Eq:ScalarSolHor}
\psi_S\sim \mathfrak{C}_S\, e^{-i\,\overline{\omega}\,\, \overline{r}_*}+\mathfrak{D}_S\, e^{+i\,\overline{\omega}\,\, \overline{r}_*}, 
\end{equation}
\noindent
which are again interpreted as the in-falling and out-going solutions, as in the vector sector \eqref{eq:incomingOutgoing}.
In order to investigate the spectral function of the scalar mesons we need to calculate the asymptotic solutions of Eq.~\eqref{Eq:ScalarSolHor}. Close to the horizon the asymptotic solutions can be written as
\noindent
\begin{equation}
    \psi_S=e^{\pm i\,\overline{\omega} \, \overline{r}_*}\left(a_{S,0}^{(\pm)}+a_{S,1}^{(\pm)}\left(u_h-u\right)+a_{S,2}^{(\pm)}\left(u_h-u\right)^2+\cdots\right)
\end{equation}
\noindent
where 
\begin{equation}
    \begin{split}
        a_{S,1}^{(\pm)}=\,&a_{S,0}^{(\pm)}\frac{u_h^2}{4(2\pm\overline{\omega} u_h i)}\left(\frac{2\left(4+\overline{k}^2\right)}{u_h}+\frac{16a_0u_h^4}{(1+u_h^4)^2}-\frac{12a_0}{1+u_h^4}+\frac{3(2+C_0^2\lambda)}{u_h^3}\right)\\
        a_{S,2}^{(\pm)}=\,&\frac{u_h^2}{16(4\pm\overline{\omega}\,u_h\,i)}\Bigg(a_{S,0}^{(\pm)}\Bigg[16-\frac{\left(56+6\overline{k}^2\right)}{u_h^2}-\frac{12C_0C_1\lambda}{u_h^3}-\frac{(30-3C_0^2\lambda)}{u_h^4}+\\
        \,& \frac{4a_0^2u_h^{2}(3-u_h^4)^2}{(1+u_h^4)^4}+\frac{4a_0(33-12u_h^2-90u_h^4-8u_h^6+5u_h^8+4u_h^{10})}{u_h(1+u_h^4)^3}\Bigg]\\
        \,&-2a_{S,1}^{(\pm)}\Bigg[\frac{2\left(4+\overline{k}^2\right)}{u_h}+\frac{42+3C_0^2\lambda}{u_h^3}\pm\frac{6\overline{\omega}\,i}{u_h^2}+\frac{4a_0(u_h^4-3)}{(1+u_h^4)^2}\Bigg]\Bigg)
    \end{split}
\end{equation}
\noindent
In turn, close to the boundary we have the asymptotic solutions identified as the normalizable, $\psi^{(1)}_S$, and non-normalizable, $\psi^{(2)}_S$, solutions:
\begin{equation}
    \begin{split}
        \psi^{(1)}_S=\,&u^{3/2}\left(b_{S,0}+b_{S,1}u+b_{S,2}u^2+b_{S,3}u^3+b_{S,4}u^4+\cdots\right)\\
        \psi^{(2)}_S=\,&u^{-1/2}\left(c_{S,0}+c_{S,1}u+c_{S,2}u^2+c_{S,3}u^3+c_{S,4}u^4+\cdots\right)+2d_S\psi^{(1)}_S\ln{[u]} \label{eq:scalar_asymptotic}
    \end{split}
\end{equation}
where the coefficients are given by 
\begin{equation} \label{eq:scalar_asymptotic_coeff}
    \begin{split}
        b_{S,1}=\,&0,\qquad b_{S,2}=\frac{b_{S,0}}{16}\left(4+2\overline{k}^2+3m_q^2\zeta^2\lambda-2\overline{\omega}^2\right),\qquad b_{S,3}=-\frac{a_0\,b_{S,0}}{10}\\
        b_{S,4}=\,&\frac{b_{S,0}}{48}\left(2+\frac{18}{u_h^4}+6m_{q}\sigma\lambda\right)+\frac{b_{S,2}}{48}\left(4+2\overline{k}^2+3m_q^2\zeta^2\lambda-2\overline{\omega}^2\right)\\
        c_{S,1}=\,& 0,\qquad c_{S,3}=-\frac{a_0\,c_{S,0}}{2},\\
        c_{S,4}=\,&\frac{c_{S,0}}{16}\left(2+\frac{2}{u_h^4}+6m_{q}\sigma\lambda\right)+\frac{c_{S,2}}{16}\left(4+2\overline{k}^2+3m_q^2\zeta^2\lambda-2\overline{\omega}^2\right)-\frac{3}{2}b_{S,2}\,d_S,\\
        d_S=\,& \frac{\left(4+2\overline{k}^2+3m_q^2\zeta^2\lambda-2\overline{\omega}^2\right)c_{S,0}}{8b_{S,0}}
    \end{split}
\end{equation}
The general solution takes the form $\psi_S = \psi^{(1)}_S + \psi^{(2)}_S$ and should contain only two independent coefficients, namely $c_{S,0}$ and $b_{S,0}$. Since terms of the form $u^{3/2 + 2n}$ which arise from the coefficient $b_{S,0}$ are identical to those arising from the coefficient $c_{S,2}$, the latter is redundant and can be set to zero.

\subsection{Analysis of the effective Schr\"odinger potentials} 
\label{sec:effective_potential}

We finish this section with an analysis of the effective potentials of the Schr\"odinger-like differential equations. The temperature dependence of the potential is useful in predicting the number of peaks in the spectral function, as we shall see in Sec.~\ref{sec:spectral_functions}. In this section, we maintain the wavenumber $k=0$ and the dilaton coupling parameter $a_0=6.5$.

\begin{figure}[ht!]
    \centering
    \includegraphics[width=0.47\textwidth]{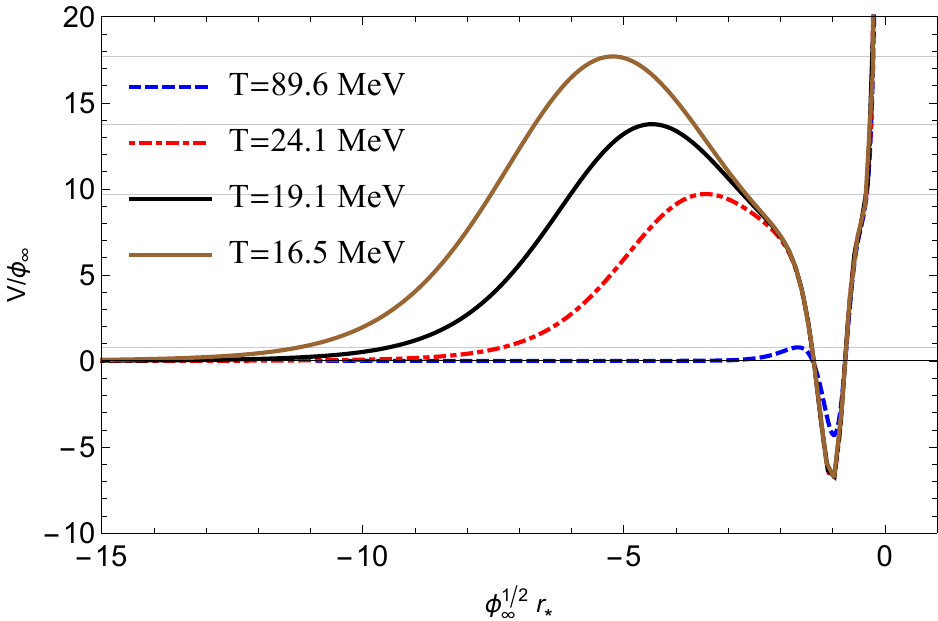}\hfill
    \includegraphics[width=0.47\textwidth]{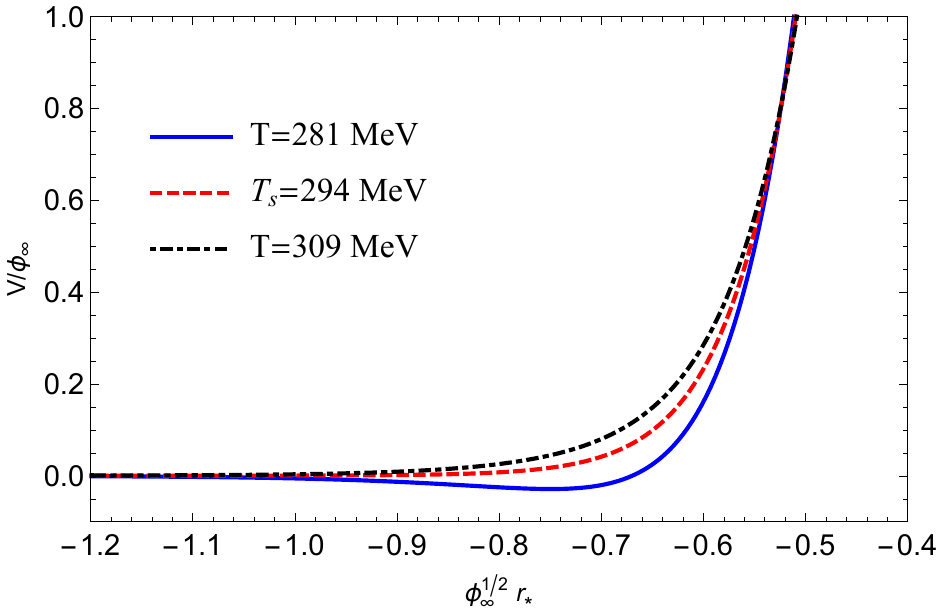}
    \caption{{\bf Left Panel}: The effective potential for the vector sector for low temperatures. The horizontal lines indicate the zero-temperature vector meson spectrum from \cite{Ballon-Bayona:2021ibm}. {\bf Right Panel}: The effective potential for high temperatures. Here, we use dilaton coupling parameter $a_0=6.5$.}
    \label{Fig:VectorPot}
\end{figure}

Let us start with the effective potential obtained in the vectorial sector \eqref{Eq:TransvPotential}. The left panel of Fig.~\ref{Fig:VectorPot} shows the potential for low temperatures. The horizontal lines represent the masses at zero-temperature, i.e., $V=m_V^2/\phi_{\infty}$, calculated in Ref.~\cite{Ballon-Bayona:2021ibm}. We calculate the temperature such that the maximum of the potential matches the zero-temperature squared mass for the first four states. This shall be our temperature of reference to observe a determined number of peaks in the spectral function. For example, for $T=16.5$ MeV, the maximum of the potential matches the squared mass of the third excited state of the zero temperature vector spectrum, as indicated by the horizontal line in the left panel of Fig.~\ref{Fig:VectorPot}. Thus, at this temperature one expects four peaks in the spectral function, which is shown in Fig.~\ref{Fig:spectra_temp}. 

Observing this figure carefully, one can affirm that there exists a temperature where the potential displays neither a maximum nor a minimum; we call this the saddle temperature $T_s$. The right panel of Fig.~\ref{Fig:VectorPot} shows the potential for $T_s=294$ MeV with dashed red line, we also show the potential for a temperature lower than $T_s$, with solid blue line, and a temperature higher than $T_s$, with dotted dashed black line. As can be seen in the figure, there is no potential well for $T\geq T_s$. 

We conclude that for very high temperatures, the potential becomes an infinite barrier on the right (where the tortoise coordinate reaches the boundary) and it goes to zero on the left (where the tortoise coordinate reaches the horizon). We then expect that bound states associated with mesons become at very high temperatures the quasinormal modes of the AdS black brane, as shown in  \cite{Miranda:2009uw,Mamani:2013ssa}. For these reasons we do not expect to clearly identify sharp peaks in the spectral function for very high temperatures. This will be confirmed in the next section, where this analysis helps us to understand, at least qualitatively, the number of expected peaks appearing in the spectral function. We should remark that it is not possible to identify the melting temperature only from the analysis of the Schr\"odinger potential. 

\begin{figure}[ht!]
    \centering
    \includegraphics[width=0.47\textwidth]{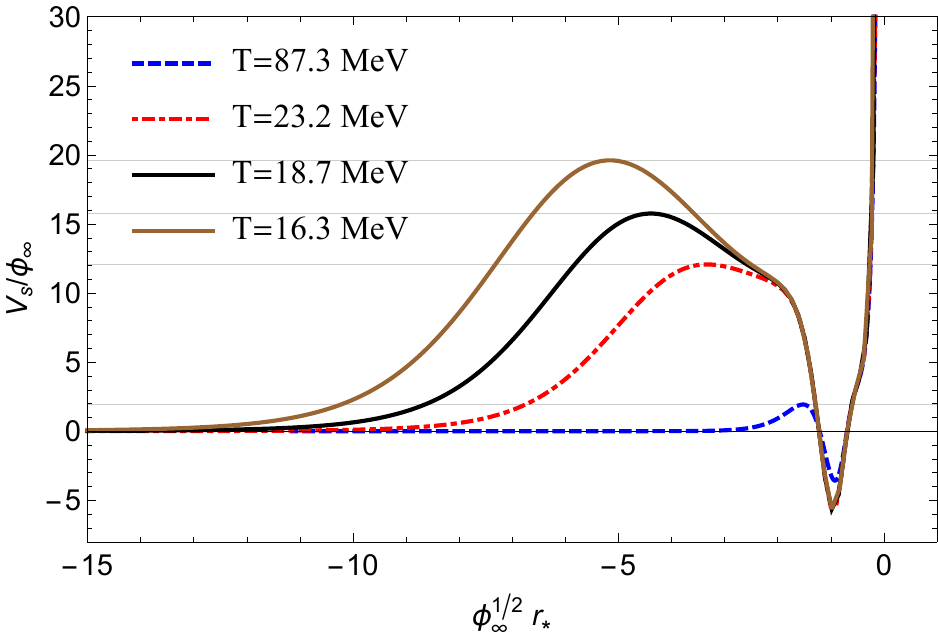}\hfill
    \includegraphics[width=0.47\textwidth]{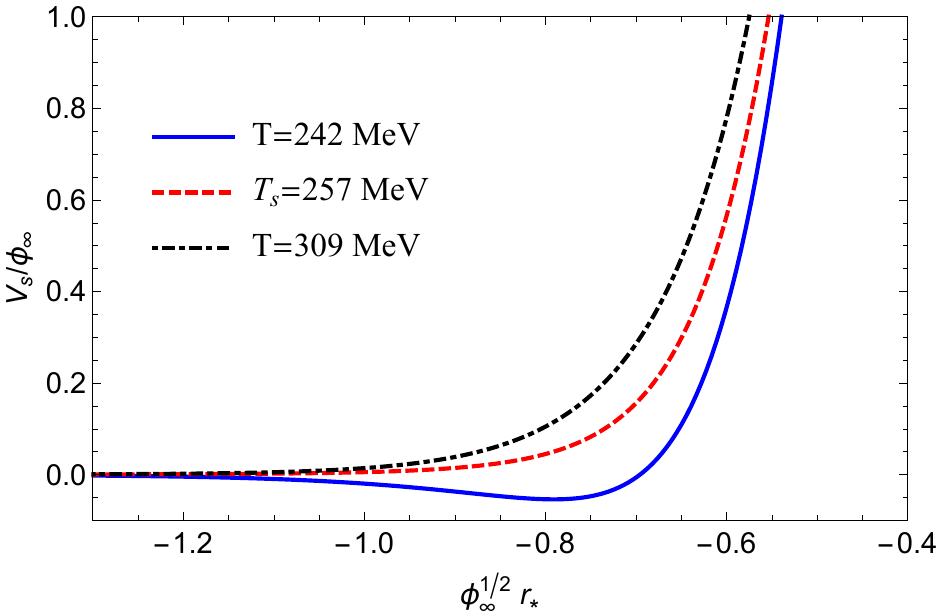}
    \caption{{\bf Left Panel}: The effective potential for the scalar sector for low temperatures. The horizontal lines indicate the zero-temperature scalar meson spectrum from \cite{Ballon-Bayona:2021ibm}. {\bf Right Panel}: The effective potential for high temperatures. Here, we use dilaton coupling parameter $a_0=6.5$.
    }
    \label{Fig:ScalarPot}
\end{figure}

We follow the same procedure for the scalar sector. A plot of the potential  \eqref{Eq:ScalarPotential_tortoise} for low temperatures regime is displayed in the left panel of Fig.~\ref{Fig:ScalarPot}. The temperatures displayed in this figure were fixed with the potential at zero temperature, i.e., $V_S=m_S^2/\phi_{\infty}$, using the results of Ref.~\cite{Ballon-Bayona:2021ibm}. As can be seen, the low- and high-temperature behavior is qualitatively similar to that of the vector sector. However, the saddle temperature for the scalar mesons is around $T_s=257$ MeV.  

In the next section we will make more precise estimates for the melting of vector and scalar mesons based on the study of the meson spectral functions.

\section{Spectral functions} \label{sec:spectral_functions}

The meson phenomenology is explored through the spectral function, which depends on frequency ($\omega$) and wavenumber ($k$). This spectral function is defined as the imaginary part of the retarded Green's function, corresponding to an incoming solution at the horizon \cite{Kovtun:2005ev}. It is calculated by finding the ratio between the contributions from the normalizable and non-normalizable asymptotic solutions, see for example \cite{Mamani:2013ssa}. The procedure is outlined in Appendix~\ref{App:Spectral_numerics}. Quasinormal modes, manifesting as sharp peaks in the spectral function, signal the presence of mesonic bound states and broaden as the temperature increases.

Bound states occur when the overlap between the normalizable solution at the boundary and the incoming solution at the horizon is large. Numerically, the spectral function is computed by solving the equations of motion for each sector at various frequencies. In this study, we focus on the scalar and vector mesons. 

For vector mesons, the spectral function is obtained by numerically solving the Schrödinger-like equation~\eqref{eq:SchrodingerVector}, starting with the asymptotic solutions at the boundary~\eqref{eq:Vector_asymptotic_boundary}, and comparing them to the incoming and outgoing solutions at the horizon~\eqref{eq:incomingOutgoing}.

For scalar mesons, a similar approach is used, though the calculation is complicated by the coupling between the scalar effective potential~\eqref{Eq:ScalarPotential_tortoise} and the tachyon. The coefficients of the asymptotic solutions~\eqref{eq:scalar_asymptotic} involve the chiral condensate $\sigma$~\eqref{eq:scalar_asymptotic_coeff}, requiring the tachyon to be solved first, as detailed in subsections \ref{SubSec:TachyonEq} and \ref{subsec:chiral_numerical}.

While our focus in this section is on the Schrödinger-like equations~\eqref{eq:SchrodingerVector} and~\eqref{Eq:ScalarSectorSchrodEq}, our results were cross-verified using the non-Schrödinger-like methods from Refs.~\cite{Colangelo:2009ra,Bartz:2016ufc}, which will be essential for analyzing the axial-vector and pseudoscalar sectors in future work.

\subsection{Scalar spectral function}
We find the spectral functions using the Son-Starinets prescription, which requires writing the on-shell action in the form
\begin{equation}
    \mathcal{S}_\mathrm{on-shell}= \left.\int \frac{d^4k}{(2\pi)^4}\phi_0(-k)\mathcal{F}(k,z)\phi_0(k)\right |_{z=0},
\end{equation}
where $\phi_0$ is the value of the field on the boundary, and the retarded Green's function is found from the flux factor $\mathcal{F}$ via the relation
\begin{equation}
    G^R(k)=-2 \mathcal{F}(z,k) \Big|_{z=0}\, .
\end{equation}

Applying this method to the scalar sector, we begin with the on-shell action 
\begin{equation}
    S_\mathrm{on-shell}= G_s \left. \int d^4x \sqrt{-g} \, e^{-a(\Phi)}g^{zz}S(z,k)\partial_z S(z,k) \right|_{z=0}.
\end{equation}
We have included a coupling constant in the scalar sector, which in our model is fixed as $G_s=1$. 
We decompose the scalar field into its source term $s_0(k)$ and the bulk-to-boundary propagator $s_k(z)$ and perform a Fourier transform 
\begin{equation}
    S_\mathrm{on-shell}= G_s \, \left.\int \frac{d^4k}{(2\pi)^4} \sqrt{-g} \, e^{-a(\Phi)}g^{zz} s_0(-k)s_{-k}(z)\partial_z s_k(z) s_0(k) \right|_{z=0}.
\end{equation}
Thus, the scalar flux factor is identified
\begin{equation}
    \mathcal{F}(z,k)= G_s e^{-a(\Phi)}\sqrt{-g}g^{zz} s_{-k}(z)\partial_z s_k(z). \label{eq:scalarGreen}
\end{equation}

To be a solution of the equation of motion \eqref{Eq:ScalarSectorEq}, the bulk-to-boundary propagator must match the in-falling condition \eqref{Eq:ScalarSolHor} at the horizon and the normalization condition $ \lim_{z \to 0} z^{\Delta - 4} s_k(z) = 1$ with $\Delta =3$ at the boundary. Thus, in terms of the Bogoliubov-transformed fields, the bulk-to-boundary propagator is written 
\begin{eqnarray}
     s_k(z)&=& e^{-B_s}\frac{\psi^-}{\mathcal{A}^-} \\
     &=& e^{-B_s} \left( \psi_S^{(2)} +\frac{\mathcal{B}^-}{\mathcal{A}^-} \psi_S^{(1)} \right) .
\end{eqnarray}
In the second line we use equation \eqref{Eq:ABcoeff} to write the incoming solution $\psi^-$ in terms of the normalizable solution $\psi_S^{(2)}$ and non-normalizable solution $\psi_S^{(1)}$   described in subsection~\ref{SubSec:scalarmesons}.  Using the near-boundary expansions \eqref{eq:scalar_asymptotic}, we insert this expression into \eqref{eq:scalarGreen}, finding 
\begin{equation}
    G^R(k) = - 2 G_s \Big [  \epsilon^{-2} + 1 + 2d_s(1+4 \log \epsilon) + 4 \frac{\mathcal{B}^-}{\mathcal{A}^-} \Big ] \, ,
\end{equation}
where $\epsilon$ is an ultraviolet regulator.

The ultraviolet divergent terms can be removed via holographic renormalization \cite{Bianchi2002}. The constant $d_S$ was defined in \eqref{eq:scalar_asymptotic_coeff}, where $b_{S,0}$ and $c_{S,0}$ are set to one.

 The spectral function is given by $\mathcal{R}(\overline{\omega},\overline{k})=2 \, \mathrm{Im} G^R(\overline{\omega},\overline{k})$, which gives
\begin{equation}
    \mathcal{R} = - 8 G_s \mathrm{Im} \frac{\mathcal{B}^-}{\mathcal{A}^-} \,.
\end{equation}
The overall factor does not affect the present analysis.

\subsection{Vector spectral function}

We follow a similar method for the spectral function for the vector sector, including both transverse and longitudinal components. 
In this case, we seek an on-shell action of the form 
\begin{equation} \label{eq:vectorOnShellFlux}
    \mathcal{S}= \left. \int \frac{d^4 k}{(2\pi)^4} V_\mu^0(-k)\mathcal{F}^{\mu \nu}(z,k) V_\nu^0(k) \right|_{z=0}\, ,
\end{equation}
with the Green functions given by 
\begin{equation}
    G^R_{\mu \nu}(k) = -2 \eta_{\mu \rho} \eta_{\nu \sigma}\mathcal{F}^{\rho \sigma}(z,k) \Big|_{z=0}
\end{equation}
For the vector field, we write the on-shell action in the radial gauge $V_{z}=0$,
\begin{equation}
  S= \frac{1}{2 g_5^2} \left. \int d^4x \,  e^{-a(\Phi)}\sqrt{-g}g^{zz}g^{\mu\nu}V_\mu\partial_z V_\nu \right|_{z=0}.
\end{equation}
We re-write the action in terms
of the Fourier transforms of the components of the vector fields
      \begin{equation}
      S=
      \frac{1}{2 g_5^2} \left. \int \frac{d\omega dq}{4\pi^2}
       e^{A_s-a(\Phi)}\Big\{V_{x^0}(z,-k)\partial_{{z}}V_{x^0}(z,k)-
      fV_j(z,-k)\partial_{{z}}V_j(z,k)\Big\}
      \right|_{z=0}\, ,
      \end{equation}
 where $j=x^1,x^2,x^3$, and summation over $j$ is implied. 
Expressing this action in terms of the gauge-invariant field \eqref{Eq:GaugeInvariantEqs} we have   
     \begin{equation}\label{eq:vector_gauge_action} 
      S=- 
     \frac{1}{2 g_5^2}  \left. \int \frac{d\omega dq}{4\pi^2}
      e^{A_s-a(\Phi)}\frac{f}{{\omega}^2}%\sum_{j}
      \left(\frac{\omega^2}{\omega^2-q^2f}\right)^{\delta_{jx^3}}
      E_{j}(z,-k)\,\partial_{{z}}\,E_{j}(z,k)\right|_{z=0}\, ,
      \end{equation}
     and $\delta_{jx^3}$ is the Kronecker delta.
We decompose the bulk field into the source term $E_{j}^0(k)$ and the bulk-to-boundary propagator  ${\cal E}_{j}(z,k)$, and the action becomes
     \begin{equation}\label{equation40} 
      S=-
      \frac{1}{2 g_5^2} \left. \int \frac{d\omega dq}{4\pi^2}
      e^{A_s-a(\Phi)}\frac{f}{{\omega}^2}
      \left(\frac{\omega^2}{\omega^2-q^2f}\right)^{\delta_{jx^3}}
      E^{0}_{j}(-k){\cal E}_{j}(z,-k)\,\partial_{z}\,{\cal E}_{j}(z,k)\,
      E^{0}_{j}(k)\right|_{z=0}\, ,
      \end{equation}
where again ${\cal E}_{j}(0,k)=1$ at the boundary
and $E_{j}(z_h,k)$ satisfies the incoming wave condition at the horizon.
Thus, the action becomes
      \begin{equation}
      \begin{split}
      & S=
      \frac{1}{2 g_5^2} \int \frac{d\omega dq}{(2\pi)^2}
      \Bigg[e^{As-a(\Phi)}f\left(\frac{\omega^2}{\omega^{2}-q^{2}f}\Big\{V^{0}_{z}(-k)\,      
      V^{0}_{z}(k)+\frac{q}{\omega} V^{0}_{x^3}(-k)V^{0}_{t}(k)+\frac{q}{\omega}V^{0}_{t}(-k)
      V^{0}_{x^3}(k)\right. \\
      &+\frac{q^2}{\omega^2} V^{0}_{t}(-k)V^{0}_{t}(k)\Big\}
      {\mathcal E}_{x^3}(z,-k)\partial_{{z}}{\mathcal E}_{x^3}(z,k)
      +\sum_{\alpha}V^{0}_{\alpha}(-k)V^{0}_{\alpha}(k){\mathcal E}_{\alpha}(z,-k)
      \partial_{{z}}{\mathcal E}_{\alpha}(z,k)
      \bigg)\Bigg]_{z=0}\, ,
      \end{split}
      \end{equation}
      where $\alpha =x^1, \, x^2$.
Comparing this to \eqref{eq:vectorOnShellFlux}, we read off the non-zero Green's function components
\begin{eqnarray}
    \frac{G^R_{t t}}{q^2}=   \frac{G^R_{x^3 x^3}}{\omega^2}= -\frac{G^R_{zt}}{q \omega} =-\frac{G^R_{tz}}{q \omega} &=& - \frac{1}{g_5^2} \frac{1}{q^2-\omega^2} \lim_{z\rightarrow 0} \frac{1}{z}\partial_z\mathcal{E}_{x^3}(z), \label{eq:vector_green1} \\
G^R_{\alpha \alpha} &=& -\frac{1}{g_5^2} \lim_{z \rightarrow 0 }\frac{1}{z}\partial_z \mathcal{E}_\alpha(z),  \label{eq:vector_green2}
\end{eqnarray}
 The expressions have been simplified by making use of the fact that $f,\, \mathcal{E}_j,$ and $e^{-a(\Phi)}$ all approach unity near the boundary $z\rightarrow 0$. 

Again, we write the gauge-invariant bulk-to-boundary propagator in terms of the in-falling solution at the boundary, given by equation \eqref{Eq:ABcoeff},
\begin{eqnarray}
    \mathcal{E}_j(z) &=& e^{-B_{T/L}} \frac{\psi^-}{\mathcal{A^-}}\\
                    &=& e^{-B_{T/L}}\left(\psi^{(2)}_j + \frac{\mathcal{B}^-}{\mathcal{A}^-} \psi^{(1)}_j\right),
\end{eqnarray}
where $B_{T/L}$ are defined in Sec. \ref{sec:vector_melting}. The longitudinal version $B_L$ is used in the case $j=x^3$, and the transverse version $B_T$ is used otherwise.

Using the near-boundary expansions \eqref{eq:Vector_asymptotic_boundary}, we insert this expression into \eqref{eq:vector_green1} and \eqref{eq:vector_green2}. We focus on the longitudinal $G^R_L(\overline{\omega},\overline{k}) \equiv G^R_{x^3x^3}(\overline{\omega},\overline{k})$ and transverse $G^R_T(\overline{\omega},\overline{k}) \equiv G^R_{\alpha \alpha}(\overline{\omega},\overline{k})$ Green's functions. 
These become 
\begin{eqnarray}
    G^R_L(\overline{\omega},\overline{k}) &=& - \frac{2}{g_5^2} \frac{\overline{\omega}^2}{\overline{\omega}^2-\overline{k}^2} \left ( 1+d_j(1+2\log \epsilon)+ \frac{\mathcal{B}^-_L}{\mathcal{A}^-_L} \right ) \, ,\\
    G^R_T(\overline{\omega},\overline{k}) &=& -  \frac{2}{g_5^2} \left ( 1 + d_j(1+2\log \epsilon)+ \frac{\mathcal{B}^-_T}{\mathcal{A}^-_T} \right ) \, ,
\end{eqnarray}
where $d_j$, was defined in \eqref{eq:Vector_coeff} where $b_{j,0}$ and $c_{j,0}$ are set to one. The longitudinal and transverse spectral functions, defined by $\mathcal{R}_{L/T}(\overline{\omega},\overline{k})=-2 \, \mathrm{Im} G^R_{L/T}(\overline{\omega},\overline{k})$, become
\begin{eqnarray}
    \mathcal{R}_L(\overline{\omega},\overline{k}) &=&  \frac{4}{g_5^2} \frac{\overline{\omega}^2}{\overline{\omega}^2-\overline{k}^2}  \mathrm{Im} \frac{\mathcal{B}^-_L}{\mathcal{A}^-_L}  \, , \\
   \mathcal{R}_{T}(\overline{\omega},\overline{k}) &=&   \frac{4}{g_5^2} \mathrm{Im}  \frac{\mathcal{B}^-_T}{\mathcal{A}^-_T}  \, . 
\end{eqnarray}
In our model the vector coupling constant is fixed as $g_5^2=12 \pi^2/N_c$, although this numerical factor  will not affect our analysis. 

\subsection{Temperature effects}

\begin{figure}[ht!]
    \centering
    \includegraphics[width=0.45\textwidth]{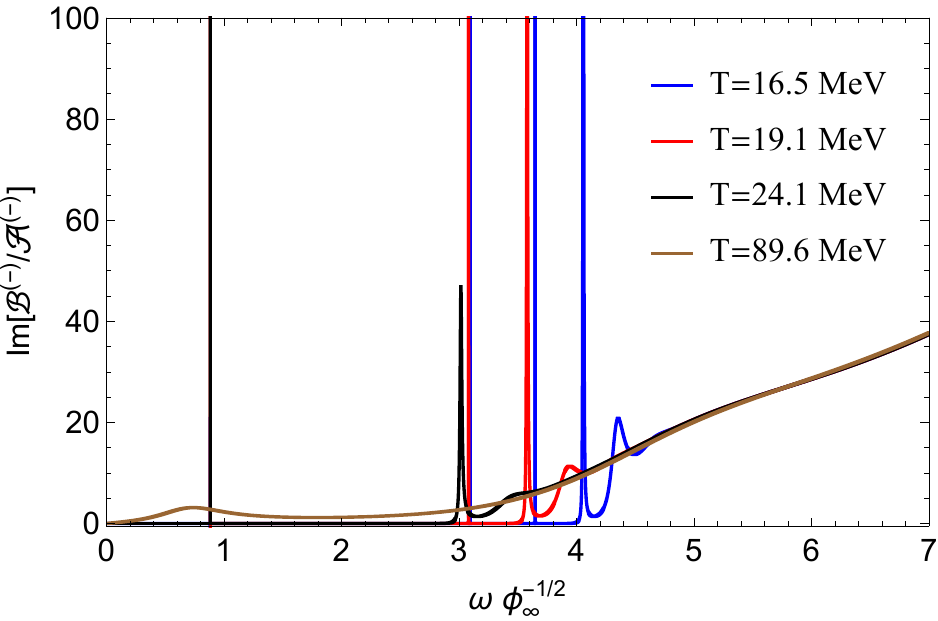}\hfill \includegraphics[width=0.45\textwidth]{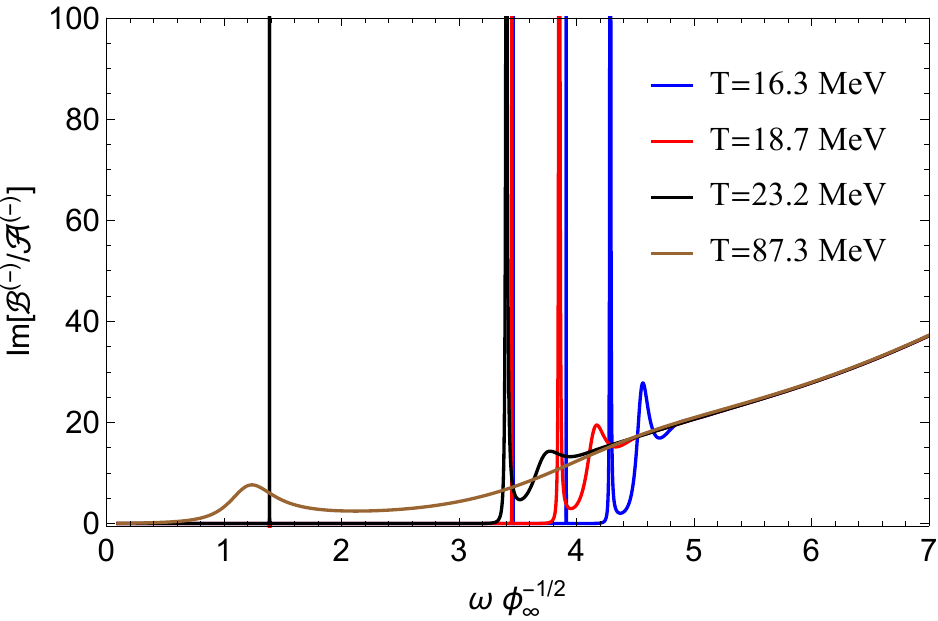}
    
    \caption{The functions for the vector (left) and scalar (right) sectors at selected values of the temperature determined  by analysis of the effective potentials, as described in Sec. \ref{sec:effective_potential}. In both sectors, the peaks corresponding to excited states shift to lower frequencies and then disappear. The shifting of the ground state is not visible at this scale for the three lowest temperatures.  }
    \label{Fig:spectra_temp}
\end{figure}

The effect of temperature on the spectral functions is shown in Fig.~\ref{Fig:spectra_temp}, at temperatures determined by the analysis of the  vector and scalar effective potentials in Sec.~\ref{sec:effective_potential}. For each sector, four narrow peaks are present at the lowest temperature, and 
these peaks broaden and shift to lower frequencies as the temperature  increases. 

The broadening and eventual disappearance of sharp peaks is interpreted as the melting of the mesons, beginning with the most highly excited states.
A meson state is considered to melt when its corresponding peak disappears from the spectral function, but
there is no definitive method to determine when the peak has vanished. In order to identify the dissociation temperature we use a Breit-Wigner fit on the first peak of the spectral function to obtain its height and width. Then, we computed the height-to-width ratio for several temperatures. We observed that this ratio develops a minimum around $T=109$ MeV for the vector mesons. We used this criterion to identify the dissociation temperature, it is expected that for temperatures higher than this value the bound states become part of the medium. It is also interesting pointing out that there are other possibilities to identify the melting temperature, one can use the inflection point of the width as in Ref.~\cite{Chen:2020afc} or investigating the structure of the complex poles of the correlation functions as implemented in Refs.~\cite{Cao:2021tcr, Cao:2020ryx}.

We would like to emphasize that the spectral function is the imaginary of the retarded correlator, which was obtained by imposing an incoming wave boundary condition at the horizon. As described in \cite{Kovtun:2005ev,Miranda:2009uw,Mamani:2013ssa}, the poles of the retarded correlator are complex and  correspond to the quasinormal frequencies of the probe fields in the AdS black brane. At low temperatures, the real and imaginary parts of the complex poles correspond to the mass and the thermal width of the quasiparticle states associated with mesons. At very high temperatures,  the mesons dissociate and the complex poles become the quasinormal modes of vector and scalar fields on the AdS black brane.

Regardless of the exact definition, all meson states in this model melt in the vicinity of $T=90-110$ MeV, which is  below the chiral transition temperature calculated in Sec.~\ref{sec:chiral} and the deconfinement temperature found by lattice methods. However, the melting temperature in this model is higher than in previous holographic meson melting models, which find all sharp peaks disappearing at temperatures of 30-60 MeV \cite{Colangelo:2009ra, Bartz:2016ufc}.

\subsection{Effects of dilaton coupling parameter \texorpdfstring{$a_0$}{}}

\begin{figure}[ht]
    \centering
    \includegraphics[width=0.45\textwidth]{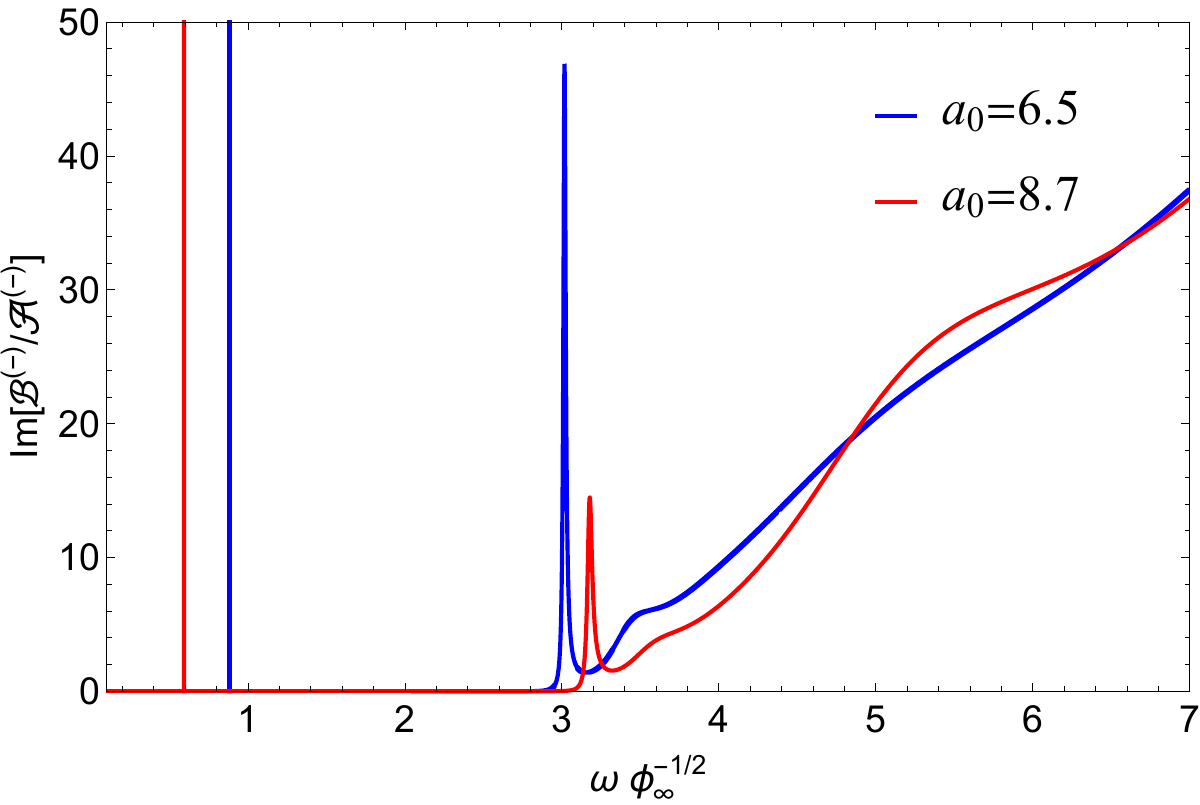}\hfill \includegraphics[width=0.45\textwidth]{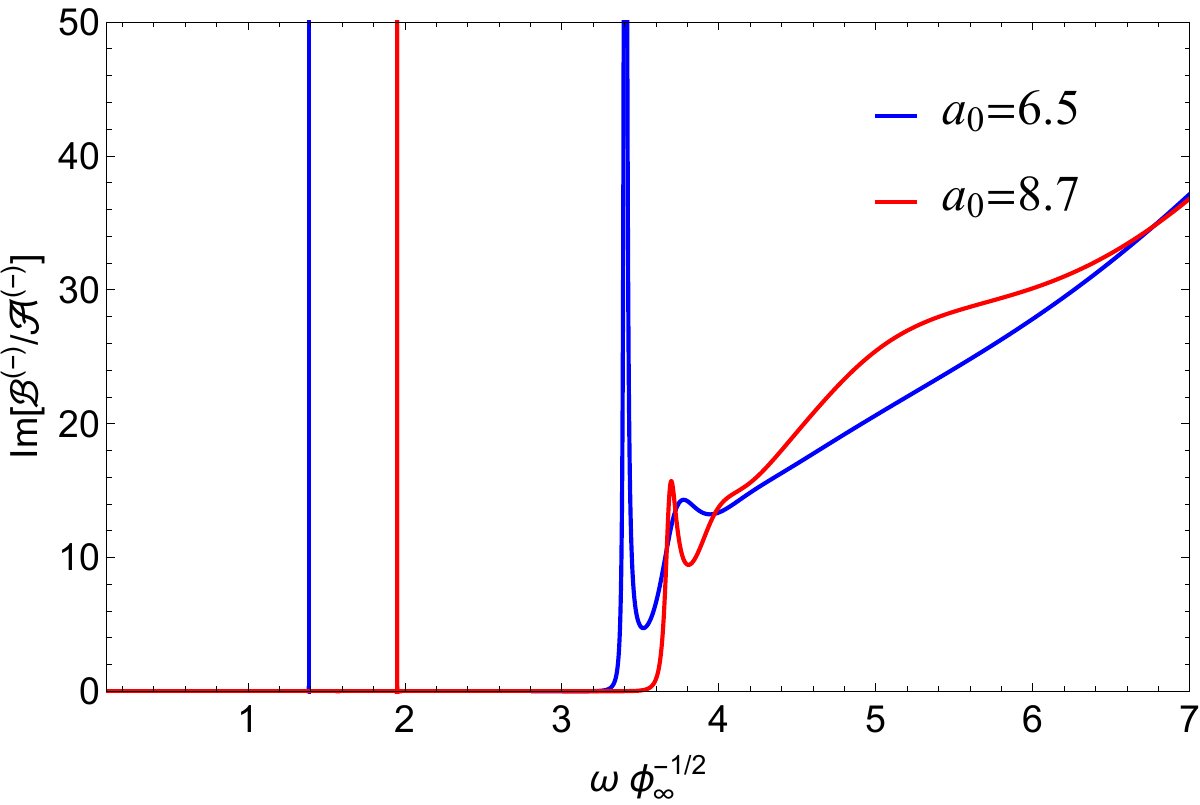}
    \caption{The spectral functions for the vector (left) and scalar (right) sectors depend the value of the dilaton coupling parameter $a_0$. Each plot shows the results for $a_0=6.5$ (blue) and $a_0=8$ (red) at the relevant temperature to produce four spectral peaks. Keeping temperature fixed, increasing $a_0$ causes all peaks to shift to higher frequencies, with the notable exception of the ground state for the vector sector, which shifts to lower frequency.}
    \label{fig:spectra_a0}
\end{figure}

The dilaton coupling parameter $a_0$ influences the zero-temperature meson mass spectrum \cite{Ballon-Bayona:2021ibm} and the critical temperature (see Fig.~\ref{Fig:Tc_vs_a0}), so it is reasonable to expect it has an effect on the spectral functions.
The vector and scalar spectral functions are shown in Fig.~\ref{fig:spectra_a0} for two values of the dilaton coupling parameter $a_0$. In each sector, we choose the temperature that produces two peaks in the spectrum, so that we can see the effect of $a_0$ on these states.

In both sectors, the excited states shift to higher frequency as $a_0$ increases. The ground states differ, as the vector ground state shifts to lower frequency but the frequency of the scalar ground state increases.
This is in agreement with the effect of $a_0$ on the zero-temperature vector meson spectrum \cite{Ballon-Bayona:2021ibm}.

Considering the ground state of the vector mesons at $T=89.6$ MeV, we perform a fit on the numerical data considering the Breit-Wigner distribution for selected values of $a_0$, observing that the width of the peak increases as $a_0$ increases. Additionally, the height-to-width ratio of the peak, which we are considering as the criteria to identify the melting temperature, increases with $a_0$. These results suggest that the melting temperature of vector mesons rises as $a_0$ increases.

\subsection{Effects of momentum \texorpdfstring{$k$}{}}

\begin{figure}[ht]
    \centering
    \includegraphics[width=0.45\textwidth]{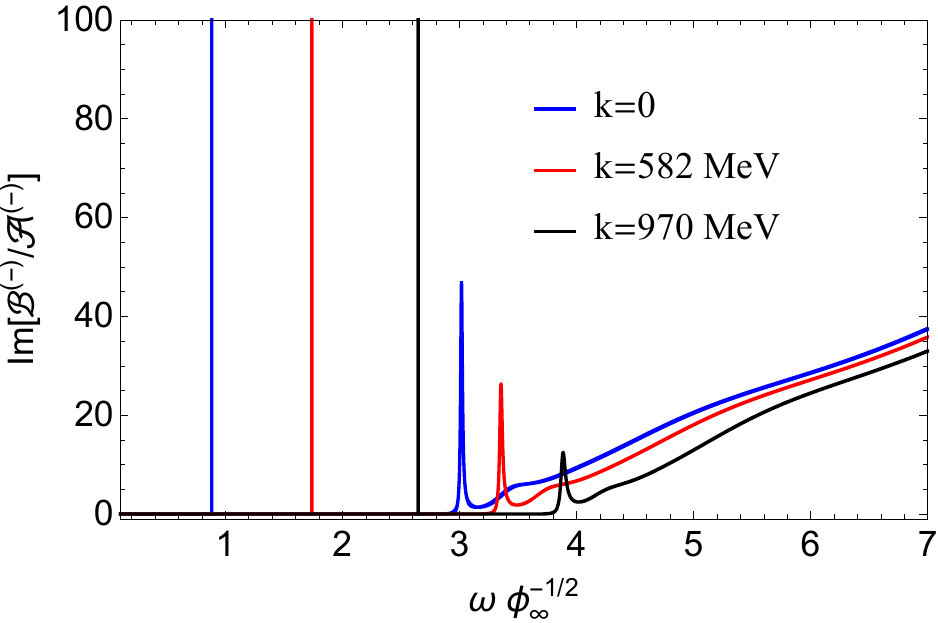}\hfill \includegraphics[width=0.45\textwidth]{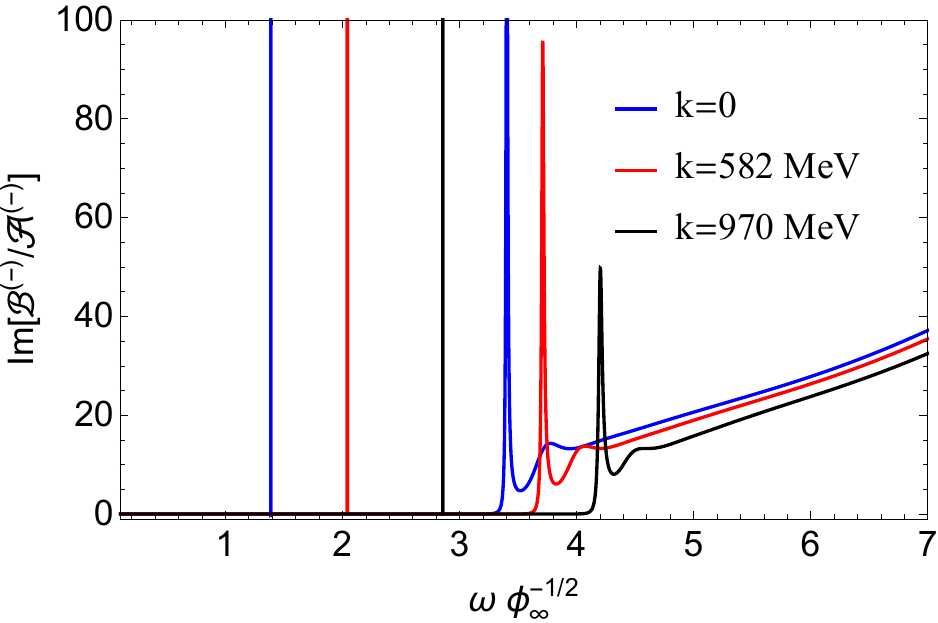}
    \caption{The spectral functions for the vector (left) and scalar (right) sectors for different values of the spatial momentum $k$. We consider $a_0=6.5$ and $T=24.1$ MeV (left) and $a_0=6.5$ and $T=23.2$ MeV (right).}
    \label{fig:spectra_ks}
\end{figure}

We also investigate the behavior of the spectral functions for different values of the wavenumber, $k$. We focus on the transverse sector, as  $k$ appears in the potential for the longitudinal sector \eqref{Eq:TransvPotential}, meaning this espression is no longer  Schr\"odinger-like for $k \neq 0$ . The results are displayed in the left panel of Fig.~\ref{fig:spectra_ks}, where solid blue line was obtained for $k=0$, solid red line was obtained for $k=582$ MeV, and solid black line was obtained for $k=970$ MeV. The shift of the peaks  to higher frequencies with increasing $k$ is evident. This means that the quasiparticle states become more energetic. This additional energy speeds up the melting process, as we can see on the second peak of the spectral function, which shortens and broadens with increasing wavenumber.

In turn, for the scalar mesons, the results are qualitatively equivalent to those obtained for the transverse sector. As shown in the right panel of Fig.~\ref{fig:spectra_ks}, increasing wavenumber shifts the peaks to higher frequencies, while also making them shorter and broader.

\section{Conclusions}
\label{Sec:Conclusions}

In this paper, we have investigated the chiral transition and the melting of vector and scalar mesons within an improved soft wall model in two-flavor holographic QCD. We started in section \ref{Sec:Model} describing the improved holographic soft wall model proposed in \cite{Ballon-Bayona:2021ibm} that leads to spontaneous chiral symmetry breaking at zero temperature. We extended the model to finite temperature introducing a 5d AdS black brane with a horizon radius $z_h$ related to the temperature $T$ of the deconfined plasma by $z_h=1/(\pi T)$. The  parameters of the model were initially fixed as in \cite{Ballon-Bayona:2021ibm} to the values $a_0=6.5$, $\phi_{\infty}=(0.388 \, {\rm MeV})^2$ and $\lambda=60$, in order to reproduce the spectrum of mesons at zero temperature. For the current quark masses $m_q = m_u = m_d$ we considered two possible scenarios: the chiral limit where $m_q=0$ and a physical quark mass of $m_q = 9 \, {\rm MeV}$.

In Section \ref{sec:chiral} we  analyzed the chiral transition at finite temperature in terms of the quark condensate. We found that the chiral transition in the chiral limit is of second-order while for finite quark mass it is a smooth crossover. We computed the corresponding critical and pseudo-critical temperatures $(T_c, T_{pc})$. We found for $a_0=6.5$ that $T_c = 128 \, {\rm MeV}$ and $T_{pc} = 129 \, {\rm MeV}$, the former result being consistent with the recent lattice QCD results for the chiral limit \cite{HotQCD:2019xnw}. We also investigated the case $a_0=8.7$ where we found $T_c = 157 \, {\rm MeV}$ and $T_{pc} = 158 \, {\rm MeV}$; the latter result being consistent with the  finite quark mass lattice QCD result of \cite{Borsanyi2020}. Moreover, we have found interesting behavior of the critical and pseudo-critical temperatures as a function of the dilaton coupling parameter $a_0$. In the chiral limit we found that $T_c$ is a monotonically increasing function of $a_0$, while at finite quark mass we found two different behaviors: for $a_0<a_{0}^{in}$, where $a_{0}^{in}$ is the inflection point, the pseudo-critical temperature decreases as we increase $a_0$, while for $a_0>a_{0}^{in}$ the pseudo-critical temperature increases as we increase $a_0$. It would be interesting to investigate the physical mechanism behind this behavior, which we leave for future research. We concluded Section~\ref{sec:chiral} by computing the quark condensate contribution to the free energy density in the chiral limit and for finite quark mass. We found that the free energy density corresponding to the non-trivial solutions are negative, which  confirm the stability of the non-trivial solutions compared with the trivial ones. For finite quark mass, we also found unphysical solutions at temperatures $T<T_{pc}$,  but these solutions have positive free energy density and therefore are not thermodynamically favored. Furthermore, in Appendix\,\ref{App:Condensate} we have included some results for the chiral condensate for $T>>T_{pc}$. At finite quark mass we have found an issue which seems to be present in all classes of interpolated softwall models reported so far in the literature, which is the fact that the chiral condensate becomes negative for very high temperatures and increases in absolute value as we increase the temperature. Finally, we have also compared our results with the results of \cite{Chelabi:2015cwn,Chelabi:2015gpc} when extended to high temperatures.

In the second part of this work, we obtained in Section \ref{Sec:mesons} the differential equations of the 5d field perturbations in the AdS black brane background associated with the propagation of vector and scalar mesons in the defonfined plasma. Considering a specific direction for the propagation, we split the vectorial sector in two parts: the longitudinal and transverse sectors. We solved the differential equation for the longitudinal sector in the hydrodynamic regime obtaining the diffusion constant of the flavor current in the deconfined plasma. We investigated the behavior of this transport coefficient as a function of the temperature for different values of the dilaton coupling parameter $a_0$, see Fig.~\ref{Fig:DT}. Next, we wrote the differential equations describing the vector and scalar mesons in the Schr\"odinger-like form in order to investigate the behavior of the effective potential at finite temperature. At low temperatures, we found that the potential displays wells characterizing the presence of quasiparticle states, see the left panel of Figs.~\ref{Fig:VectorPot} and \ref{Fig:ScalarPot}. However, as the temperature increases, the potential wells disappear characterizing the melting of the quasiparticle states, as shown in the right panel of Figs.~\ref{Fig:VectorPot} and \ref{Fig:ScalarPot}. 

To confirm the observations of Section \ref{Sec:mesons}, we computed the spectral functions in Section~\ref{sec:spectral_functions}. First, we investigated the spectral functions for selected values of the temperature, observing that the number of peaks decreases with the increasing of the temperature, characterizing the melting process, see Fig.~\ref{Fig:spectra_temp}. For the model parameters fixed as  $a_0=6.5$, $\phi_{\infty}=(0.388 \, {\rm MeV})^2$ and $\lambda=60$, in order to reproduce the spectrum of mesons at zero temperature, we found that the melting temperature for vector and scalar mesons lies within the range of $90 - 110 \, {\rm MeV}$, which is close to but lower than the pseudocritical temperature for the chiral transition $T_{pc} = 129 \, {\rm MeV}$.  
We also investigated how sensitive the spectral functions are to the dilaton coupling parameter, $a_0$. In this case, we obtained different results for the vector and scalar mesons. The first peak of the vector mesons shifts to low frequencies with the increasing of $a_0$ (state becomes less energetic), while the first peak of the scalar mesons shifts to high frequencies with the increasing of $a_0$ (state becomes more energetic), see Fig.~\ref{fig:spectra_a0}. These results suggest that increasing the dilaton coupling parameter, the melting temperature for the vector mesons increases while the melting temperature for the scalar mesons decreases. 
Finally, we found that the melting process is also affected by the wavenumber, which shift the peaks of the spectral functions to the right and speeds up the melting process.

A possible extension of this work is solving the axial-vector and pseudo-scalar mesons. The analysis would determine the melting process of the pion, and if a finite-temperature analogue to the Gell-Mann--Oakes--Renner (GOR) relation can be defined.  It would also be interesting to investigate meson melting in the holographic QCD model proposed in \cite{Ballon-Bayona:2023zal} where confinement and spontaneous chiral symmetry breaking are realized using a 5d background based on Einstein-dilaton gravity. Another possible line of research is the study of effects of a strong magnetic field in the chiral transition in the spirit of \cite{Ballon-Bayona:2020xtf,Bohra:2020qom}.

\section*{Acknowledgments}

The work of the author A.B-B is partially funded by Conselho Nacional de Desenvolvimento Cient\'\i fico e Tecnol\'ogico (CNPq, Brazil), Grant No. 314000/2021-6, and Coordena\c{c}\~ao de Aperfei\c{c}oamento do Pessoal de N\'ivel Superior (CAPES, Brazil), Finance Code 001.

\appendix

\section{Chiral condensate: high temperatures}
\label{App:Condensate}
\indent

\begin{figure}[htb]
\centering
\includegraphics[scale=0.37]{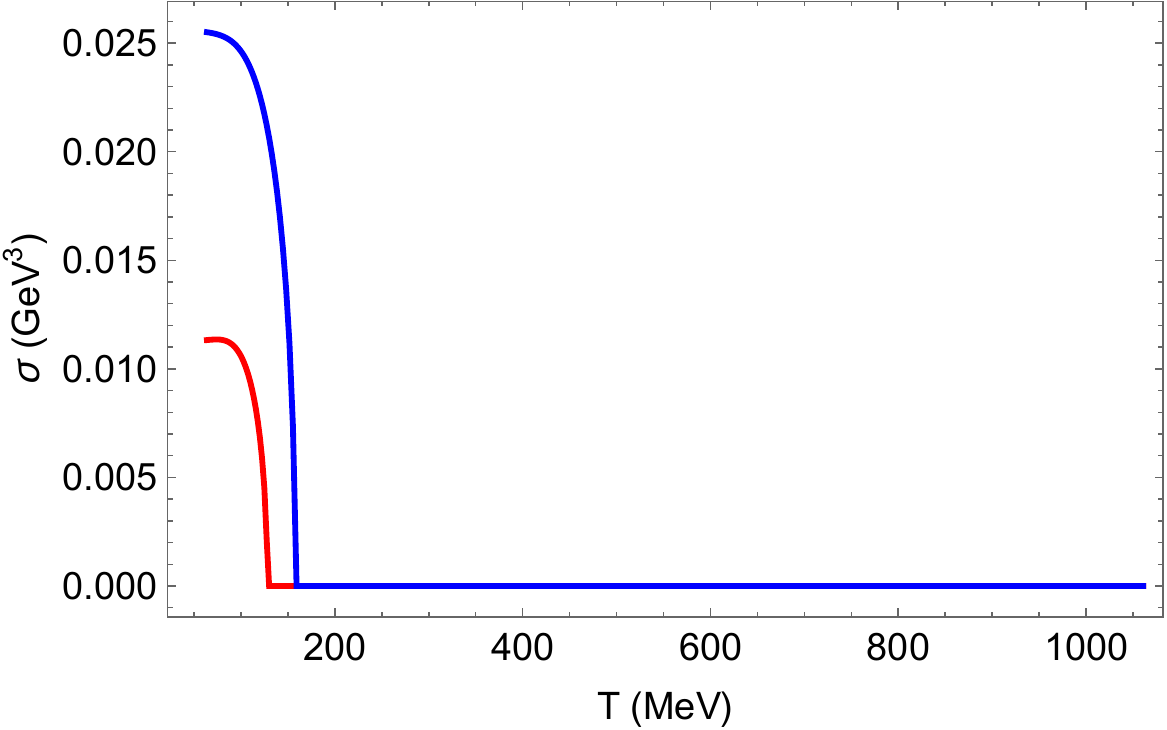}
\hfill
\includegraphics[scale=0.37]{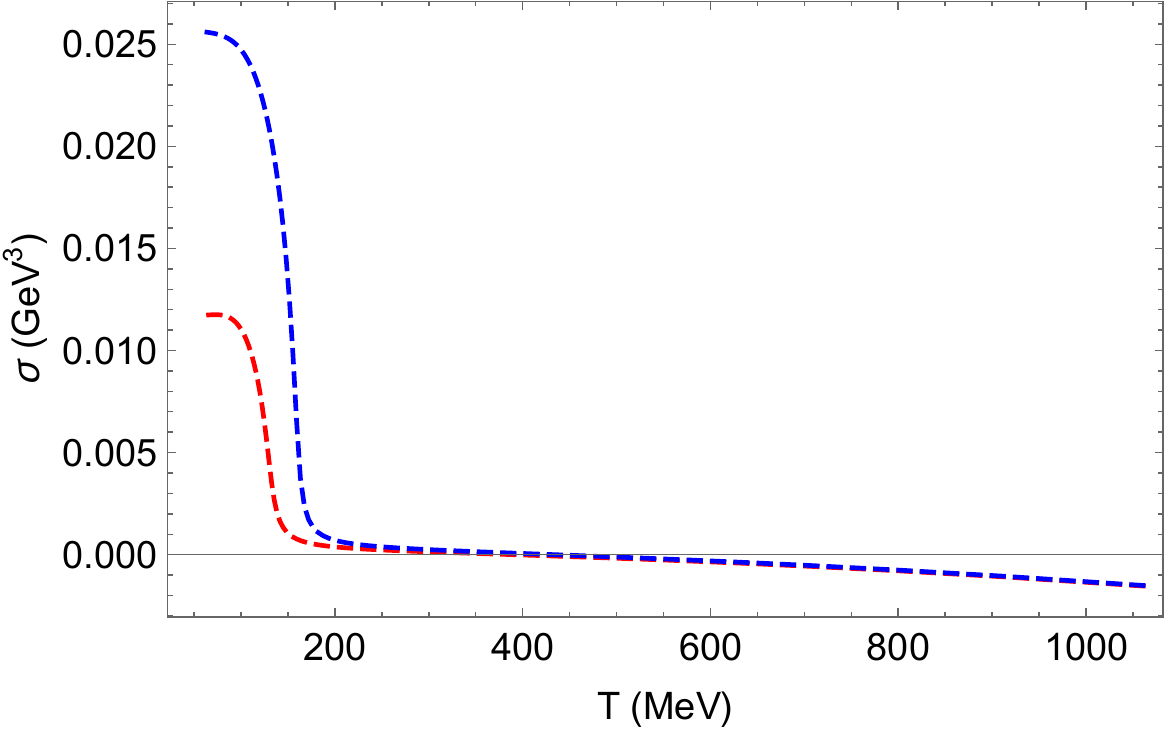}
    \caption{\textbf{Left Panel}: Chiral condensate as a function of the temperature in the chiral limit ($m_{q}=0$) for $a_0=6.5$ (red curve) and $a_0=8.7$ (blue curve) extended to high temperature.  \textbf{Right Panel}: Chiral condensate as a function of the temperature for finite quark mass ($m_q = 9$ MeV)  for $a_0=6.5$ (dashed red curve) and $a_0=8.7$ (dashed blue curve) extended to high temperature.}
\label{Fig:ChiralTransitionHighT}
\end{figure}

In this Appendix we comment on the behavior of the chiral condensate as a function of the temperature extended to higher temperatures. The result, displayed in Fig.\ref{Fig:ChiralTransitionHighT}, shows that at finite quark mass the chiral condensate becomes negative after a characteristic temperature, $T_0$, and its absolute value increases as we approach very high temperatures (see right panel of Fig.\ref{Fig:ChiralTransitionHighT}). In the chiral limit on the other hand this behavior does not occur, and the chiral condensate vanishes for $T>T_c$ (see left panel of Fig.\ref{Fig:ChiralTransitionHighT}).

For finite quark mass, we define the characteristic temperature, $T_0$ such that $\sigma(T_0)=0$, and for $T>T_0$ the chiral condensate $\sigma$ increases in absolute value as we increase the temperature. This temperature depends on the value of $a_0$ and increases as we increase $a_0$. For $a_0 = 6.5$ we find $T_0 = 390.17$ MeV and for $a_0 = 8.7$ we find $T_0 = 434$ MeV.

\begin{figure}[htb]
\centering
\includegraphics[scale=0.7]{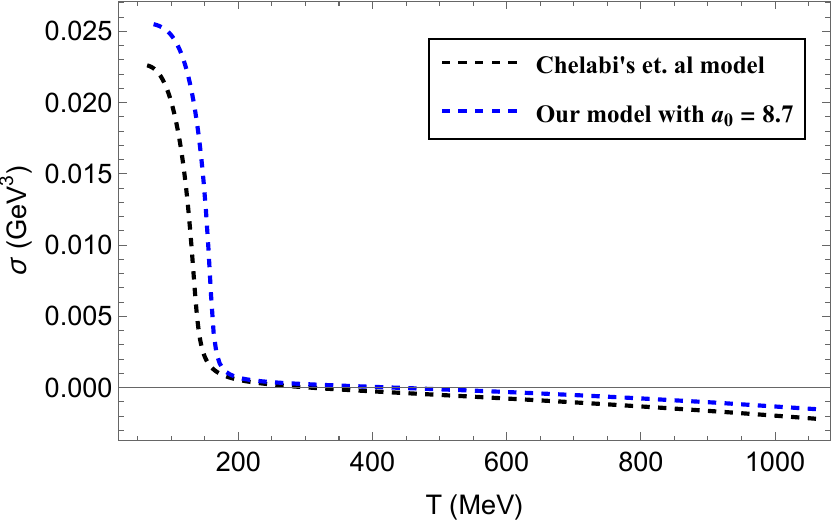}
\caption{Comparison between the chiral transition at finite quark mass ($m_q=9$ MeV) in our model with $a_0=8.7$ and in the model of Refs.\cite{Chelabi:2015cwn,Chelabi:2015gpc}. Note that both softwall interpolations suffer from the same issue, i.e, the chiral condensate going negative for $T>T_0$, where $T_0$ is the characteristic temperature  previously defined.}
\label{Fig:Comparison}
\end{figure}

This weird behavior of the chiral condensate at finite quark mass for high temperatures is not particular to our interpolation for the dilaton coupling. We have also checked that the first-proposed dilaton interpolation of \cite{Chelabi:2015cwn,Chelabi:2015gpc} has also the same issue, as shown in Fig.\ref{Fig:Comparison}, where we compare our results for the chiral condensate as a function of the temperature at finite quark mass with $a_0=8.7$ and the results of \cite{Chelabi:2015cwn,Chelabi:2015gpc}. 

\section{Hydrodynamic solution of the transverse sector}
\label{App:Hydro}

For the transverse sector we follow the same procedure we did for the longitudinal sector. First, considering the transformation 
\noindent
\begin{equation}
    E_{\alpha}=f^{-\frac{i \mathfrak{w}}{4}}F_{\alpha}(\mathfrak{u}).
\end{equation}
\noindent
Plugging in the differential equation \eqref{Eq:TransVectorN2}, we get the differential equation for $F_{\alpha}(\mathfrak{u})$
\noindent
\begin{equation}
    \begin{split}
    F_{\alpha}''+&\Big(A_s'-a'+\frac{(2-i\lambda\mathfrak{w})f'}{2f}\Big)F_{\alpha}'\\
    &+\lambda\Big(\frac{\lambda\mathfrak{w}^2(f'^{\,2}-16)}{16f^2}+\frac{4\mathfrak{q}^2\lambda+i\mathfrak{w}As'f'-i\mathfrak{w}a'f'+i\mathfrak{w}f''}{4f}\Big)F_{\alpha}=0.
    \end{split}
\end{equation}
\noindent
Next, considering the expansion, $F_{\alpha}(\mathfrak{u})=F_{\alpha,0}(\mathfrak{u})+\lambda F_{\alpha,1}(\mathfrak{u})+\lambda^2 F_{\alpha,2}(\mathfrak{u})+\dots$. By plugging this expansion in the last equation one get differential equations for $F_{\alpha,0}$, $F_{\alpha,1}$ and so on. However, here we are going to consider just up to first order in $\lambda$
\noindent
\begin{align}
    &F_{\alpha,0}''+\Big(A_s'-a'+\frac{f'}{f}\Big)F_{\alpha,0}'=0 \label{Eq:TransF0}\\
    &F_{\alpha,1}''+\Big(A_s'-a'+\frac{f'}{f}\Big)F_{\alpha,1}'-\frac{i\mathfrak{w}f'}{2f}F_{\alpha,0}'-i\mathfrak{w}\Big(\frac{A_s'f'-a'f'+f''}{4f}\Big)F_{\alpha,0}=0. \label{Eq:TransF1}
\end{align}
\noindent
The regular solution for $\eqref{Eq:TransF0}$ is $F_{\alpha,0}(\mathfrak{u})=F_{\alpha,0}=\text{constant}$.
\noindent
\begin{equation}
    F_{\alpha,1}=C_2+\frac{i\mathfrak{w}}{4}F_{\alpha,0}\ln{f(u)}+C_1\int_{0}^{u}\frac{e^{-A_s(x)+a(x)}}{f(x)}dx
\end{equation}
\noindent
\begin{align}
    C_1=\,&-i\frac{\mathfrak{w}F_{\alpha,0}f'(\mathfrak{u}_h)e^{A_s(\mathfrak{u}_h)-a(\mathfrak{u}_h)}}{4}=-i\frac{F_{\alpha,0}\mathfrak{w}}{4}\lim_{y\to\mathfrak{u}_h}\frac{\ln{f(y)}}{\int_{0}^{y}\frac{e^{-A_s(x)+a(x)}}{f(x)}dx}\\
    C_2=\,&0
\end{align}
\noindent
Then, the complete solution, up to first-order in $\lambda$, is
\begin{equation}
    E_{\alpha}=f^{-\frac{i\mathfrak{w}}{4}}\left(F_{\alpha,0}+\lambda F_{\alpha,1}(u)\right)
\end{equation}
\noindent
Expanding close to the boundary one get
\noindent
\begin{equation}
    E_{\alpha}=F_0-\frac{i\,F_0\mathfrak{w}\mathfrak{u}_h^2}{2}e^{-a(\mathfrak{u}_h)}u^2+\cdots
\end{equation}
\noindent
By imposing Dirichlet boundary condition, $E_{\alpha}(0)=0$, we do not get any solution compatible with the hydrodynamic regime.

\section{Spectral functions: Numerical method}
\label{App:Spectral_numerics}

Here we write additional details on the method we used to calculate the spectral functions, we follow the ideas of Refs.~\cite{Miranda:2009uw, Mamani:2013ssa,Bartz:2016ufc, Mamani:2018uxf}. In the following analysis we consider the incoming, $\Theta^{(-)}_j$, and outgoing, $\Theta^{(+)}_j$, solutions at the horizon, we also consider the non-normalizable, $\Theta^{(2)}_j$, and normalizable, $\Theta^{(1)}_j$, solutions close to the boundary. It is worth pointing out that the subscript $j$ in the above functions refers to $j=V$ the vector sector, and $j=S$ the scalar sector. In the following analysis, $\Theta_{V}$ can be the solution of the differential equation \eqref{Eq:TransVectorN2}, or the solution of the Schr\"odinger-like equation \eqref{eq:SchrodingerVector}. The same is valid for the scalar sector, $\Theta_{S}$ can be the solution of the differential equation \eqref{Eq:ScalarSectorEq}, or the solution of the Schr\"odinger-like equation \eqref{Eq:ScalarSectorSchrodEq}. This means that one can use the solutions original equations to calculate the spectral functions or the solutions of the Schr\"odinger-like equations to get the spectral functions. We start by writing the incoming, $\Theta^{(-)}_j$, and outgoing, $\Theta^{(+)}_j$, solutions at the horizon as a linear combination of the non-normalizable, $\Theta^{(2)}_j$, and normalizable, $\Theta^{(1)}_j$, solutions close to the boundary.
\noindent
\begin{subequations}\label{Eq:ABcoeff}
\begin{align}
\Theta^{(-)}_j=\,& \mathcal{A}^{(-)}_j\,\Theta^{(2)}_j+\mathcal{B}^{(-)}_j\,\Theta^{(1)}_{j},\\
\Theta^{(+)}_j=\,& \mathcal{A}^{(+)}_j\,\Theta^{(2)}_j+\mathcal{B}^{(+)}_j\,\Theta^{(1)}_{j},
\end{align}
\end{subequations}
\noindent
where the subscript $j$ indicates the meson sector we are considering. One can rewrite this  system of equations in  matrix form,
\noindent
\begin{equation}\label{Eq:MatrixCoeffN1}
\begin{pmatrix}
\Theta^{(-)}_j\\
\Theta^{(+)}_j
\end{pmatrix}
=
\begin{pmatrix}
\mathcal{A}^{(-)}_j & \mathcal{B}^{(-)}_j\\
\mathcal{A}^{(+)}_j & \mathcal{B}^{(+)}_j
\end{pmatrix}
\begin{pmatrix}
\Theta^{(2)}_j\\
\Theta^{(1)}_j
\end{pmatrix}.
\end{equation}
\noindent

On the other hand, one can write the non-normalizable and normalizable solutions as a linear combination of the incoming and outgoing solutions:
\noindent
\begin{subequations}\label{eq:BoundHori}
\begin{align}
\Theta^{(2)}_j=\,& \mathcal{C}^{(2)}_j\,\Theta^{(-)}_j+\mathcal{D}^{(2)}_j\,\Theta^{(+)}_{j},\\
\Theta^{(1)}_j=\,& \mathcal{C}^{(1)}_j\,\Theta^{(-)}_j+\mathcal{D}^{(1)}_j\,\Theta^{(+)}_{j},
\end{align}
\end{subequations}
\noindent
which can be written in matrix form, 
\noindent
\begin{equation}\label{Eq:MatrixCoeffN2}
\begin{pmatrix}
\Theta^{(2)}_j\\
\Theta^{(1)}_j
\end{pmatrix}
=
\begin{pmatrix}
\mathcal{C}^{(2)}_j & \mathcal{D}^{(2)}_j\\
\mathcal{C}^{(2)}_j & \mathcal{D}^{(1)}_j
\end{pmatrix}
\begin{pmatrix}
\Theta^{(-)}_j\\
\Theta^{(+)}_j
\end{pmatrix}.
\end{equation}
\noindent
By comparing equations \eqref{Eq:MatrixCoeffN1} and \eqref{Eq:MatrixCoeffN2}, we get the following relation between the coefficients:
\noindent
\begin{equation}\label{Eq:MatrixCoeff}
\begin{pmatrix}
\mathcal{A}^{(-)}_j & \mathcal{B}^{(-)}_j\\
\mathcal{A}^{(+)}_j & \mathcal{B}^{(+)}_j
\end{pmatrix}
=
\begin{pmatrix}
\mathcal{C}^{(2)}_j & \mathcal{D}^{(2)}_j\\
\mathcal{C}^{(1)}_j & \mathcal{D}^{(1)}_j
\end{pmatrix}^{-1}.
\end{equation}
\noindent
We want the incoming solutions, related to retarded Green functions. Thus, the coefficients $\mathcal{A}^{(-)}_j$ and $\mathcal{B}^{(-)}_j$ in terms of the coefficients $\mathcal{C}_j$ and $\mathcal{D}_j$ are:
\noindent
\begin{equation*}
    \mathcal{A}^{(-)}_j=\frac{\mathcal{D}^{(1)}_j}{\mathcal{C}^{(2)}_j\mathcal{D}^{(1)}_j-\mathcal{C}^{(1)}_j\mathcal{D}^{(2)}_j},\qquad  \mathcal{B}^{(-)}_j=-\frac{\mathcal{D}^{(2)}_j}{\mathcal{C}^{(2)}_j\mathcal{D}^{(1)}_j-\mathcal{C}^{(1)}_j\mathcal{D}^{(2)}_j},
\end{equation*}
\noindent
the ratio is given by
\noindent
\begin{equation}\label{Eq:ABDCoeff}
    \frac{\mathcal{B}^{(-)}_j}{\mathcal{A}^{(-)}_j}=-\frac{\mathcal{D}^{(2)}_j}{\mathcal{D}^{(1)}_j}.
\end{equation}
\noindent

Next, the idea is to express the coefficients $\mathcal{D}^{(2)}_j$ and $\mathcal{D}^{(1)}_j$ in terms of the solutions to the differential equation, the Schr\"odinger-like or the original differential equation. Then, to calculate these coefficients one can rewrite Eqs.~\eqref{eq:BoundHori} and their derivatives in a compact form:
\noindent
    \begin{align}
    \Theta^{(n)}_j=\,& \mathcal{C}^{(n)}_j\,\Theta^{(-)}_j+\mathcal{D}^{(n)}_j\,\Theta^{(+)}_{j},\qquad n=1,2\\
    \partial_z\Theta^{(n)}_j=\,& \mathcal{C}^{(n)}_j\,\partial_z\Theta^{(-)}_j+\mathcal{D}^{(n)}_j\,\partial_z\Theta^{(+)}_{j},\qquad n=1,2
    \end{align}
    \noindent
which can be written in the matrix form
\begin{equation}
    \begin{pmatrix}
        \Theta^{(n)}_j\\
        \partial_z\Theta^{(n)}_j
    \end{pmatrix}
    =\begin{pmatrix}
        \Theta^{(-)}_j & \Theta^{(+)}_{j}\\
        \partial_z\Theta^{(-)}_j & \partial_z\Theta^{(+)}_{j}
    \end{pmatrix}
    \begin{pmatrix}
        \mathcal{C}^{(n)}_j \\
        \mathcal{D}^{(n)}_j
    \end{pmatrix}.
    \qquad n=1,2
\end{equation}
Then, inverting the matrix one gets
\begin{equation}
    \begin{pmatrix}
        \mathcal{C}^{(n)}_j \\
        \mathcal{D}^{(n)}_j
    \end{pmatrix}
    =\begin{pmatrix}
        \Theta^{(-)}_j & \Theta^{(+)}_{j}\\
        \partial_z\Theta^{(-)}_j & \partial_z\Theta^{(+)}_{j}
    \end{pmatrix}^{-1}
    \begin{pmatrix}
        \Theta^{(n)}_j\\
        \partial_z\Theta^{(n)}_j
    \end{pmatrix}. \qquad n=1,2
\end{equation}
After some algebraic manipulations we finally get the coefficients
\begin{align}
    \mathcal{C}^{(n)}_j=\,& \frac{\Theta^{(+)}_j\partial_z\Theta^{(n)}_j - \partial_z\Theta^{(+)}_{j}\Theta^{(n)}_j }{\Theta^{(+)}_{j}\partial_z\Theta^{(-)}_j - \Theta^{(-)}_j \partial_z\Theta^{(+)}_{j}},\qquad \mathcal{D}^{(n)}_j=\frac{\partial_z\Theta^{(-)}_j \Theta^{(n)}_j-\Theta^{(-)}_j \partial_z\Theta^{(n)}_j}{\Theta^{(+)}_{j}\partial_z\Theta^{(-)}_j - \Theta^{(-)}_j \partial_z\Theta^{(+)}_{j}}.
\end{align}
In this way, one gets the ratio
\begin{equation}\label{Eq:DCoeff}
    \frac{\mathcal{D}^{(2)}_j}{\mathcal{D}^{(1)}_j}=\frac{\partial_z\Theta^{(-)}_j \Theta^{(2)}_j-\Theta^{(-)}_j \partial_z\Theta^{(2)}_j}{\partial_z\Theta^{(-)}_j \Theta^{(1)}_j-\Theta^{(-)}_j \partial_z\Theta^{(1)}_j}=\frac{\Theta^{(-)}_j \partial_z\Theta^{(2)}_j-\partial_z\Theta^{(-)}_j \Theta^{(2)}_j}{\Theta^{(-)}_j \partial_z\Theta^{(1)}_j-\partial_z\Theta^{(-)}_j \Theta^{(1)}_j}.
\end{equation}
\noindent
The analysis above is an indirect way to compute the ratio of $\mathcal{B}^{(-)}_j/\mathcal{A}^{(-)}_j$, first computing the ratio of $\mathcal{D}^{(2)}_j/\mathcal{D}^{(1)}_j$, through Eq.~\eqref{Eq:ABDCoeff}. However, as a check of consistency, we now calculate the ratio of $\mathcal{B}^{(-)}_j/\mathcal{A}^{(-)}_j$ directly from Eqs.~\eqref{Eq:ABcoeff}. Writing Eqs.~\eqref{Eq:ABcoeff} and their derivatives in a compact form:
\noindent
\begin{subequations}
\begin{align}
\Theta^{(m)}_j=\,& \mathcal{A}^{(m)}_j\,\Theta^{(2)}_j+\mathcal{B}^{(m)}_j\,\Theta^{(1)}_{j},\qquad m=-,+\\
\partial_z\Theta^{(m)}_j=\,& \mathcal{A}^{(m)}_j\,\partial_z\Theta^{(2)}_j+\mathcal{B}^{(m)}_j\,\partial_z\Theta^{(1)}_{j}, \qquad m=-,+
\end{align}
\end{subequations}
which can be written in matrix form
\begin{equation}
    \begin{pmatrix}
        \Theta^{(m)}_j    \\
        \partial_z\Theta^{(m)}_j
    \end{pmatrix}
    =\begin{pmatrix}
        \Theta^{(2)}_j & \Theta^{(1)}_{j} \\
        \partial_z\Theta^{(2)}_j & \partial_z\Theta^{(1)}_{j}
    \end{pmatrix}
    \begin{pmatrix}
        \mathcal{A}^{(m)}_j \\
        \mathcal{B}^{(m)}_j
    \end{pmatrix}.
    \qquad m=-,+
\end{equation}
Inverting the matrix to get the coefficients $\mathcal{A}_j^{(m)}$ and $\mathcal{B}_j^{(m)}$, 
\noindent
\begin{equation}
    \begin{pmatrix}
        \mathcal{A}^{(m)}_j \\
        \mathcal{B}^{(m)}_j
    \end{pmatrix}
    =
    \begin{pmatrix}
        \Theta^{(2)}_j & \Theta^{(1)}_{j} \\
        \partial_z\Theta^{(2)}_j & \partial_z\Theta^{(1)}_{j}
    \end{pmatrix}^{-1}
        \begin{pmatrix}
        \Theta^{(m)}_j    \\
        \partial_z\Theta^{(m)}_j
    \end{pmatrix}.
\end{equation}
\noindent
Thus, one gets
\begin{equation}\label{Eq:ABcoeffN2}
    \mathcal{A}^{(m)}_j=\frac{\Theta^{(m)}_j\partial_z\Theta^{(1)}_{j}-\Theta^{(1)}_{j}\partial_z\Theta^{(m)}_j}{\Theta^{(2)}_{j}\partial_z\Theta^{(1)}_{j}-\Theta^{(1)}_{j}\partial_z\Theta^{(2)}_{j}},\qquad \mathcal{B}^{(m)}_j=-\frac{\Theta^{(m)}_j\partial_z\Theta^{(2)}_{j}-\Theta^{(2)}_{j}\partial_z\Theta^{(m)}_j}{\Theta^{(2)}_{j}\partial_z\Theta^{(1)}_{j}-\Theta^{(1)}_{j}\partial_z\Theta^{(2)}_{j}}.
\end{equation}
Now, we calculate the ratio considering $m=-$
\begin{equation}\label{Eq:RatioBA}
    \frac{\mathcal{B}^{(-)}_j}{\mathcal{A}^{(-)}_j}=-\frac{\Theta^{(-)}_j\partial_z\Theta^{(2)}_{j}-\partial_z\Theta^{(-)}_j\Theta^{(2)}_{j}}{\Theta^{(-)}_j\partial_z\Theta^{(1)}_{j}-\partial_z\Theta^{(-)}_j\Theta^{(1)}_{j}}.
\end{equation}
\noindent
which is the same result obtained using the indirect way, i.e., \eqref{Eq:DCoeff}. In order to get the spectral function one needs to solve numerically the differential equations. The way we proceed was to integrate the differential equations starting from the boundary, i.e., we defined a cutoff close to the boundary. Then, we used the asymptotic solution close to the boundary as boundary condition, finally evaluating the numerical solution close to the horizon. In this way, we get the solutions for $\Theta_{j}^{(1)}$ and $\Theta_{j}^{(2)}$. Then, we get the spectral function plugging these numerical solutions, and the asymptotic solutions close to the horizon $\Theta_{j}^{(-)}$ in Eq.~\eqref{Eq:RatioBA}.

\bibliographystyle{utphys}

\bibliography{HologXSBFinT_v2}

\end{document}